\documentclass[pre,onecolumn,superscriptaddress,floatfix,raggedbottom]{revtex4-2}
\usepackage{bm,mathrsfs}
\usepackage{graphicx}
\usepackage{amsmath,amssymb,amsgen,amsbsy,bm}
\usepackage{xcolor}
\usepackage{etoolbox}
\usepackage{ulem}

\newcommand{\DV}[1]{{\color{black} #1}}

\newcommand{\Wi}{\mathrm{Wi}}
\newcommand{\Wicr}{\mathrm{Wi}_{\rm cr}}

\begin{document}

\title{Polymer stretching in laminar and random flows: entropic characterization}

\author{Stefano Musacchio}
\affiliation{Dipartimento di Fisica and INFN, Università di Torino, Via P. Giuria 1, 10125 Torino, Italy}

\author{Victor Steinberg}
\affiliation{Department of Physics of Complex Systems, Weizmann Institute of Science, Rehovot 76100, Israel}

\author{Dario Vincenzi}%\thanks{Also Associate, International Centre for Theoretical Sciences, Tata Institute of Fundamental Research,Bangalore 560089, India}
\affiliation{Universit\'e C\^ote d'Azur, CNRS, LJAD, 06100 Nice, France}

\date{\today}

\begin{abstract}
Polymers in non-uniform flows undergo strong deformation, which in the presence of persistent stretching can result in the 
coil-stretch transition. 
\textcolor{black}{
This phenomenon has been characterized by using the formalism of nonequilibrium statistical mechanics. In particular,
the entropy of the polymer extension reaches a maximum at the transition.
We extend the entropic characterization of the
coil-stretch transition by studying the differential entropy of the polymer fractional extension} in
a set of laminar and random velocity fields that are benchmarks for the study of polymer stretching in flow. 
In the case of random velocity fields, a suitable description of the transition is obtained by considering
the entropy of the logarithm of the extension instead of the entropy of the extension itself. 
Entropy emerges as an effective tool for capturing the coil-stretch transition and comparing its
features in different flows.
\end{abstract}
 
%%%%%%%%%%%%%%%%%%%%%%%
%\pacs{...}

\maketitle

\section{Introduction}

The configuration of a \textcolor{black}{linear} polymer in a moving fluid
drastically changes from coiled to fully stretched 
when the Weissenberg number Wi, \textit{i.e.}~the
product of the
characteristic velocity gradient and the polymer relaxation 
time, exceeds a critical threshold. This phenomenon is known as the coil-stretch
transition \cite{deGennes} and is observed in both laminar \cite{psc97,sbsc03} 
and random flows \cite{gcs05,ls10,ls14},
even though with partially different features in the two cases.
\textcolor{black}{In addition to the nature of the flow, the coil-stretch transition is
influenced by whether the polymer is confined spatially or not \cite{ttd10}, as well as by
the solvent quality \cite{ssp10,ru12, ru13}, the polymer 
concentration \cite{prabhakar}, and the occurence of knots along the polymer \cite{snkd18}.
Moreover, a coil-stretch transition has also been observed 
in ring polymers \cite{delatorre2005,rings,hss16}, entangled polymer melts \cite{sek18b,sek18,kki18}, and elastic-sheets \cite{yg21}.}

\textcolor{black}{Characterizing the coil-stretch transition and accurately identifying the value of
Wi at which it occurs is essential for predicting the viscoelastic properties of 
polymer solutions. For instance, phenomena such as turbulent drag reduction \cite{g14,bc18,x19} and elastic turbulence \cite{steinberg21,steinberg22,datta22} are observed
only if the polymers get sufficiently stretched by the flow. 
In the case of linear polymers,}
several observables have been used to characterize the coil-stretch transition.
A natural quantity is the steady-state distribution of polymer extensions \cite{psc97,sbsc03,gcs05,ls10,ls14}, which changes 
dramatically near to the critical Wi:
the mean increases rapidly,
the coefficient of variation attains its maximum value, and
the peak shifts from the equilibrium extension $R_{\rm eq}$ to
the maximum length $L$ (here $R_{\rm eq}$ is 
the polymer root mean square extension in the absence of flow).
Another characterization
considers the equilibration time of the statistics of polymer extension \cite{cpv06,gs08}
or alternatively the autocorrelation time of the extension \cite{wg10}.
Near the coil-stretch transition these properties are strongly amplified,
and this causes a critical slowing down of the stretching dynamics.

{\color{black}The coil-stretch transition has also been studied using
non-equilibrium thermodynamic concepts \cite{tccp07,ls13,latinwo14,latinwo2014nonequilibrium,vtc15,gc16,gc18}.
In particular, it has been shown that, in an extensional flow,
the entropy of the polymer extension is maximum at the critical Wi \cite{latinwo2014nonequilibrium}.
This result has a clear interpretation in terms of information, since the 
entropy quantifies the ``randomness'' of the extension within an ensemble of 
polymers.
In the coiled and stretched states the information concerning the polymer elongation reaches a maximum
because the distribution of polymer extensions is peaked 
around a single value ($R_{\rm eq}$ and $L$, respectively). 
These states are hence minima of entropy. 
Conversely, the broadening of the probability distribution of polymer elongations at the transition 
corresponds to a loss of information and therefore a maximum of entropy. 
Recently, Sultanov \textit{et al.}~\cite{ssflls21} have extended this result to random flows
by measuring the entropy of the extension of an ensemble of T4 DNA molecules of maximum length $L=71.7\mu m$ and radius of gyration $R_g=1.5\mu m$
in an elastic turbulence of von K\'arm\'an flow \cite{gs00,steinberg21}.}

Here we pursue the entropic characterization of the coil-stretch transition
by examining a set of analytical and numerical flows.
\textcolor{black}{The goal of our study is to show that} since it concentrates information on the statistics of  
the extension in a single scalar quantity, 
entropy is as an effective tool for comparing polymer stretching in different flows.

\section{Polymer model and flow configurations}
\label{sect:models}

The polymer is modelled as a finitely extensible nonlinear elastic
(FENE) dumbbell  \cite{bird,l88,g18}. \DV{In the Appendix, we show that, 
for a bead-spring chain, the results only differ in minor quantitative details.}
The evolution equation for the
polymer end-to-end vector $\mathbf{R}$ is 
\begin{equation}
\frac{d\mathbf{R}}{dt}=\boldsymbol{\kappa}(t)\cdot\mathbf{R}-f(R)\,
\frac{\mathbf{R}}{2\tau}
+\sqrt{\frac{R_0^2}{\tau}}\,\boldsymbol{\xi}(t),
\label{eq:dumbbell}
\end{equation}
where $\kappa_{ij}(t)=\nabla_j u_i(t)$ is the velocity gradient at the centre of
mass of the polymer, $\tau$ is the polymer longest relaxation time, $R_0=R_{\rm eq}/\sqrt{3}$, 
$f(R)=(1-R^2/L^2)^{-1}$, and $\bm\xi(t)$ is three-dimensional white noise.
Within this model, the radius of gyration is $R_g = R_{\rm eq}/2 = \frac{\sqrt{3}}{2} R_0$
and the extensibility parameter is defined as $b=(L/R_0)^2$ \cite{bird}.
The dumbbell model can in principle be refined to include effects such as hydrodynamic
interactions or a conformation-dependent drag force \cite{bird,l88}. 
Given that our work is focused on the entropic characterization of the coil-stretch transition,
rather than on the properties of the dumbbell model itself, for the sake of simplicity 
we restrict to the basic version of the model,  
which in any case has proved useful for a qualitative, and sometimes 
even quantitative, understanding of the coil-stretch transition,
in both steady \cite{deGennes,psc97,lpsc97,sbsc03} and random \cite{l73,bfl00,c00,bc18} flows.

Calculating the entropy requires obtaining the stationary probability density function (PDF) 
of the extension, $P(R)$, from Eq.~\eqref{eq:dumbbell}, analytically or numerically.
We shall consider the following set of model flows, which have been widely employed in
the study of polymer stretching and are representative
of more complex situations.
\DV{For each of these flows, the main results on the statistics of the extension are recalled below.}

\paragraph*{Extensional flow}
The uniaxial extensional flow
$\mathbf{u}=\gamma(-x/2,-y/2,z)$ is the first 
configuration in which the coil-stretch transition has been predicted \cite{deGennes} and
observed experimentally \cite{psc97}. It consists of a direction of pure stretching and
two directions of compression with magnitudes that ensure incompressibility.
The Weissenberg number is defined as $\mathrm{Wi}=\gamma\tau$ and
its critical value is $\Wicr=1/2$.
If the rescaled end-to-end vector $\bm\rho=\mathbf{R}/L$ is expressed
in spherical coordinates as 
$\bm{\rho}=\rho(\sin\theta\cos\phi,\sin\theta\sin\phi,\cos\theta)$, then
the stationary PDF of $\bm\rho$ is
\begin{equation}
P(\bm\rho) \propto \left(1-\rho^2\right)^{b/2}
\exp\left\{\frac{b\mathrm{Wi}}{2}\rho^2[3\cos^2(\theta)-1]\right\},
\end{equation}
where $b=(L/R_0)^2$ is the extensibility parameter \cite{bird}. An integration over the angular variables yields
\begin{equation}
P(\rho)= 2\pi \rho^2\int_0^\pi P(\bm\rho)\sin\theta\,d\theta
\propto \rho \,e^{-\frac{b\Wi}{2}\rho^2} \left(1-\rho^2\right)^{b/2}
\operatorname{erf}\left(\mathrm{i}\,\sqrt{\frac{3 b\mathrm{Wi}}{2}}\,\rho\right),
\end{equation}
where erf is the error function. 

\paragraph*{Shear flow}
In a linear shear flow $\mathbf{u}=(\sigma y,0,0)$, the coil-stretch transition is not observed \cite{sbc99}. 
Owing to thermal fluctuations, the dynamics of
the polymer indeed consists of a sequence of tumbling events which in turn correspond to as many coiling and stretching events,
so that persistent stretching is never realized \cite{stsc05a,tbsc05,gs06}.
Nevertheless, it will be instructive to study the entropy of polymer extension also in this configuration and compare its behaviour with that
observed in other flows. The Weisseinberg number is $\Wi=\sigma\tau$, and the PDF of $\rho$
is now calculated numerically by means of Brownian Dynamics simulations of Eq.~\eqref{eq:dumbbell},
where the nonlinearity of the elastic force is resolved by using \"Ottinger's rejection algorithm \cite{o96}.

\paragraph*{Batchelor-Kraichnan (BK) flow}
In random flows, it is convenient to define the Weissenberg number as $\Wi=\lambda\tau$,
where $\lambda$ is the Lyapunov exponent of the flow, \textit{i.e.}~the average stretching rate of line elements.
A general theory of the coil-stretch transition in random flows has been developed by Balkovsky \textit{et al.}~\cite{bfl00}
for linear polymer elasticity (Oldroyd-B model) and by Chertkov \cite{c00} for nonlinear polymer elasticity (FENE model).
For intermediate extensions $1/\sqrt{b} \ll \rho\ll 1$, the PDF of $\rho$ behaves as $\rho^{-1-\alpha}$ 
with $\alpha$ decreasing as a function of Wi and crossing zero at $\Wi=1/2$.
Therefore, in the limit $L\to\infty$ the PDF of $\rho$ is not normalizable
if $\Wi\geqslant 1/2$. This is interpreted as an indication that the coil-stretch transition
also exists in random flows and the critical Wi is again $\Wicr=1/2$.
For finite $L$, the measured slope may be affected by the nonlinearity of the 
elastic force, but the theory still implies an analogous strong modification of 
$P(R)$ at $\Wi_c$ \cite{c00}.

The BK flow has been used extensively in the analytical study of
turbulent transport below the viscous-dissipation scale
(see Ref.~\cite{fgv01} and, for applications to polymer dynamics, Ref.~\cite{pav16} and 
references therein).
The velocity gradient is an isotropic
tensorial white noise with correlation
$\langle \kappa_{ij}(t)\kappa_{kl}(t')\rangle=
\lambda\delta(t-t')(4\delta_{ik}\delta_{jl}-\delta_{ij}\delta_{kl}-\delta_{il}\delta_{jk})/3$,
where $i,j=1,2,3$.
The properties of this stochastic flow allow an exact calculation of $P(\rho)$ (see Refs.~\cite{c00,mav05}):
\begin{equation}
P(\rho) = c\, \rho^2\left(1+\frac{2b\mathrm{Wi}}{3}\,\rho^2\right)^{-\beta}\left(1-\rho^2\right)^\beta
\end{equation}
with $\mathrm{Wi}=\lambda\tau$, $\beta^{-1}=2(b^{-1}+2\mathrm{Wi}/3)$, and
\begin{equation}
\begin{split}
c^{-1}=&\frac{\sqrt{\pi}\,\Gamma(\beta+1)}{4\Gamma(5/2+\beta)}\,
_2 F_1(3/2,\beta;3/2+\beta+1;-2b\mathrm{Wi}/3).
\end{split}
\end{equation}
Here $\Gamma$ and $_2 F_1$ denote the Gamma and hypergeometric functions, respectively.
In this case, 
the exponent of the power-law region of the PDF is
$\alpha=2\beta-3\approx -3(1-1/2\Wi)$ for $b\gg 1$.

\paragraph*{Isotropic turbulence}
Although useful for a qualitative study of the coil-stretch transition,
the BK flow is Gaussian and has zero correlation time. It therefore cannot capture all
features of a fully turbulent flow.
Thus, we also consider polymers in homogeneous isotropic turbulence.
To this end, we use a database of
Lagrangian trajectories from a direct numerical simulation (DNS) of
\textcolor{black}{the three-dimensional incompressible Navier-Stokes equations \cite{jr17,vwrp21}.
These  were solved by means of a standard, fully de-aliased pseudo-spectral method 
on a cubic domain of size $2\pi$ with $512^3$ collocation points and periodic boundary conditions.
The flow was driven to a stationary state by an external force that maintained a constant energy
injection rate. The choice of  the kinematic viscosity and the energy injection rate yielded 
a Taylor-microscale Reynolds number $R_\lambda=111$. 
The velocity gradient $\boldsymbol{\kappa}(t)$
was calculated along a large number of fluid trajectories by using a bilinear interpolation algorithm.
Here we use this database of time series of $\boldsymbol{\kappa}(t)$ 
to solve Eq.~\eqref{eq:dumbbell} for an ensemble of $10^4$ polymers.
\textcolor{black}{Since attention is restricted to single polymer dynamics, the polymer feedback on the flow is disregarded.}
Equation~\eqref{eq:dumbbell}} is again solved by using \"Ottinger's rejection algorithm \cite{o96}.
The values of the parameters of the dumbbell model in the DNS are $R_0=1$ and $L=18$. 
The extensibility parameter is $b=(L/R_0)^2 = 18^2$. 
The effect of thermal noise on the position of the centre of mass is disregarded, since thermal fluctuations
are negligible compared to the fluctuations of the turbulent velocity field.
The Weissenberg number is again defined in terms of the Lyapunov exponent. 
\textcolor{black}{In the present simulation $\lambda\approx 0.136\tau_K$, where $\tau_K$ is
the Kolmogorov dissipation time scale, in accordance with previous estimates in isotropic turbulence \cite{bbbcmt06,bm17}.}
Numerical simulations of isotropic turbulence \cite{bcm03,wg10} have shown that
$P(\rho)$ behaves as a power of $\rho$ for $1/\sqrt{b}\ll \rho\ll 1$, as predicted by
Balkovksy \textit{et al.} \cite{bfl00}. 

{\color{black}
\section{Differential entropy}
Recent studies~\cite{latinwo2014nonequilibrium,ssflls21} have shown that the 
$P(R)$ broadens at the transition from the coiled to the stretched state. 
The broadening of $P(R)$ can be interpreted as a loss of information concerning the polymer elongation, 
which can be quantified in terms of the Shannon entropy.
The results of Refs.~\cite{latinwo2014nonequilibrium,ssflls21} confirms that the Shannon entropy attains a maximum  
at the Weissenberg number corresponding to the transition.
%Here we extend that the idea of the entropic characterization
%for the coil-stretch transition to a broader class of flows.

The definition of the information entropy in the case of polymers requires some care, 
because the elongation $R$ is a dimensioned and continuous variable.
The extension of the discrete information entropy
to a continuous random variable $x \in \mathbb{R}$
has been originally proposed by Shannon,
which introduced the concept of \textit{differential} entropy:  
\begin{equation}
\label{eq:diff-entropy}
S_x = - \int P(x)\, \log[P(x)]\, dx, 
\end{equation}
where $P(x)$ is the steady-state probability density function of the variable $x$.  
Unlike the discrete entropy, the differential entropy has some drawbacks.
It can be negative (because $P(x)$ can assume values larger than $1$)
and it is not invariant under a change of variable.
Considering a trasformation $y = g(x)$, the corresponding entropies are related via
$S_y \le S_x + \int P(x)\, \log \left\vert {\frac {\partial g}{\partial x}}\right\vert dx$,   
where $\left\vert {\frac {\partial g}{\partial x}}\right\vert$ is the Jacobian of the transformation $g$
and the equality holds if the transform is a bijection.
A further problem arises if $x$ is a dimensioned variable, as in the case of the polymer elongation.
In this case, $P(x)$ has the dimension of $[1/x]$ and therefore the defintion of $\log[P(x)]$ is inappropriate.

Jaynes proposed to fix these issues by introducing the concept of {\it relative} entropy
in terms of the limiting density of discrete points~\cite{Jaynes63}.
The relative entropy is defined as the negative Kullback-Leibler divergence~\cite{kullback195110}
from the distribution $P$ to the reference invariant measure $M$: 
\begin{equation}
\label{eq:diff-entropy-KL}
H_x = - D_{KL}(P||M) = - \int P(x)\, \log\left[ \frac{P(x)}{M(x)}\right]\, dx.  
\end{equation}
An alternative approach consists in non-dimensionalizing the argument of the logarithm
with a characteristic scale $\Delta$ homogeneous to $x$:  
\begin{equation}
\label{eq:diff-entropy-delta}
S^\Delta_x = - \int P(x)\, \log[P(x)\Delta]\, dx.   
\end{equation}
The modified entropy $S^\Delta_x$ is related to the Shannon differential
entropy via $S^\Delta_x = S_x -\log{\Delta}$. 
Recalling that $P(x)\Delta = P(x/\Delta)$,
the modified entropy is equivalent to the differential entropy of the
dimensionless quantity $x/\Delta$, that is, $S^\Delta_x = S_{x/\Delta}$.

For the specific case of the polymer end-to-end distance $R$,
we propose here to use the differential entropy of the
rescaled polymer elongation $\rho=|\mathbf{R}|/L$,
defined as\begin{equation}
\label{eq:entropy-R}
S_\rho=-\int_0^1 P(\rho)\, \ln[P(\rho)]\, d\rho.
\end{equation}
Here and in the following we use the natural logarithm
in the definition of the entrophy.  
The variable $\rho$ assumes values in the interval $(0,1)$
and its PDF is normalized as follows: $\int_0^1 P(\rho)\,d\rho=1$.
The differential entropy~(\ref{eq:entropy-R})
is equivalent to the relative entropy~(\ref{eq:diff-entropy-KL})
with respect to the uniform measure $M(\rho) = 1$. 
As a consequence,
the entropy $S_\rho$ assumes only negative or null values.  
The maximum value $S_\rho=0$ is attained
for a uniform distribution $P(\rho)=1$,
which corresponds to the maximun uncertainty of the polymer elongation. 
The differential entropy~(\ref{eq:entropy-R}) is also equivalent
to the modified entropy~(\ref{eq:diff-entropy-delta}) of $R$ with characteristic scale $L$: 
$S_\rho = S^L_R = -\int P(R)\ln[P(R)L] dR$.
The relation $S_\rho = S^L_R = S_R -\ln{L}$ shows that a change of the maximum elongation $L$
corresponds to a shift of the entrophy.   
In the next Section we will discuss how this property can be exploited
to determine the parameter $L$ of the dumbbell model which best fits the experimental data.
For the cases of random or turbulent flows, we will also show that
the coil-stretch transition is better described
in terms of the differential entrophy of
the dimensionless quantity $y=\ln \rho$.
The entropies of $y$ and $\rho$ are related via
$S_y = S_\rho - \mathbb{E}[\ln(\rho)]$. 
Finally, we note that in the case of a dumbbell in potential flows,
the differential entropy $S_R = S_\rho + \ln L$ coincides with the
thermodynamic entropy in~\cite{latinwo2014nonequilibrium}.
}

\section{Results}

\begin{figure}[!t]
\centering
\includegraphics[width=0.485\columnwidth]{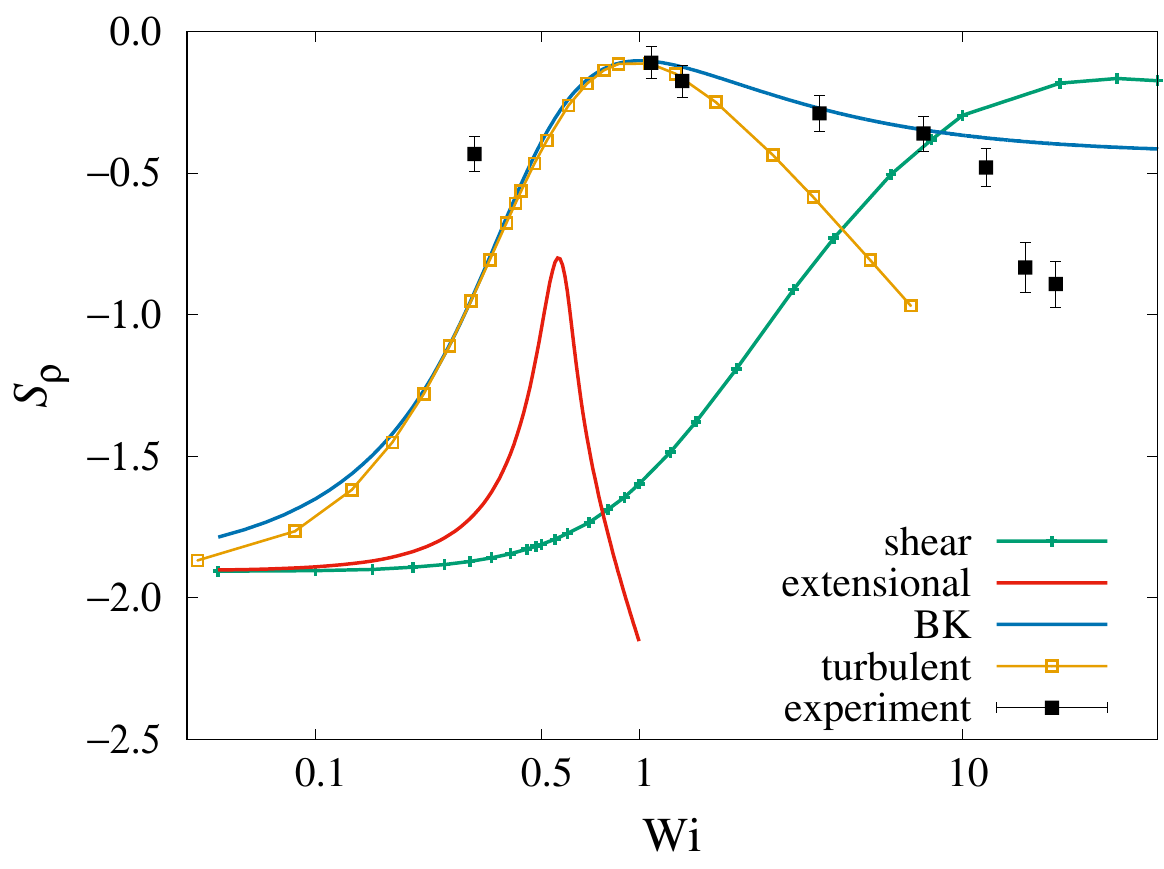}%
\hfill%
\includegraphics[width=0.505\columnwidth]{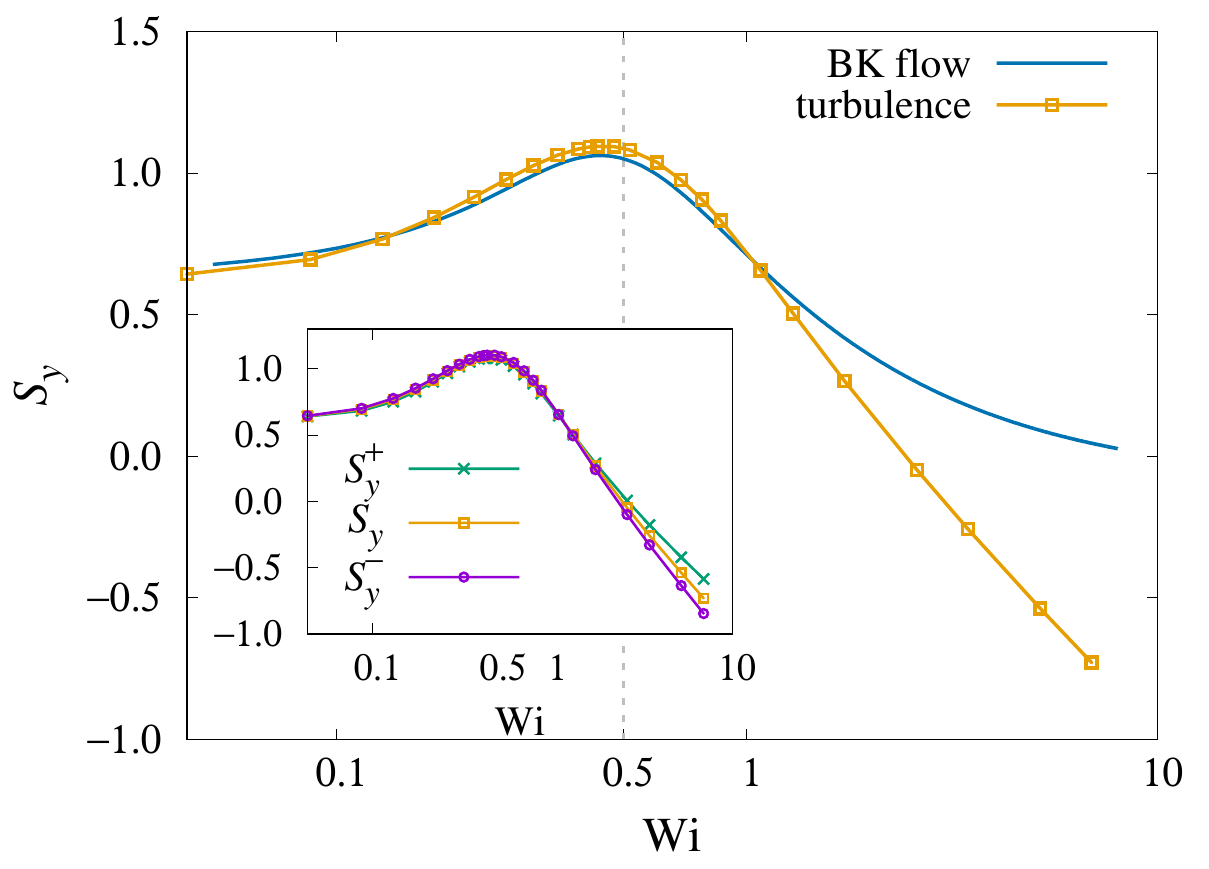}
\caption{Left: Entropy of $\rho$ vs Wi for different flows.
In all cases (except for the experimental data) 
the extensibility parameter is set to $b=18^2$. 
The experimental data have been translated vertically 
by $\Delta S_\rho=0.33$, which corresponds to a fit to a dumbbell
with $b\approx 30^2$.
Right: Entropy of $y=\ln\rho$ for the BK and turbulent flows and for
the same parameters as in the left panel.
\textcolor{black}{The inset shows $S_y$ together with the entropies $S^-_y$ and $S_y^+$
associated with the PDFs of $y$ conditional on $Q<0$ and $Q>0$, respectively.}
}
\label{fig:entropy}
\end{figure}

\textcolor{black}{
The differential entropy of the rescaled polymer length is now used to characterize and compare the coil-stretch transition
in the flows introduced in Sect.~\ref{sect:models}.
$S_\rho$ is plotted in Fig.~\ref{fig:entropy} (left panel)
as a function of Wi;} in all cases (except for the experimental data) 
the extensibility parameter is set to a representative value of $b =18^2$.

In the extensional flow, 
$S_\rho$ displays a narrow maximum at Wi near critical, \textit{i.e.}~the coil-stretch transition is marked by a strong amplification of the entropy of $\rho$. 
This behaviour reflects the fact that, at both small and large Wi, the PDF of $\rho$ is dominated by
a peak (near to either $1/\sqrt{b}$ or $1$), whereas only in a narrow range of Wi around $\Wicr$ the PDF has a broader 
shape. A large variety of polymer configurations is thus observed at the coil-stretch transition, as can be appreciated  by direct inspection 
of the time series of $\rho$ \cite{psc99,gs08}.

In the shear flow, 
$S_\rho$ starts growing in an appreciable way only when Wi is significantly greater than $\Wicr$. 
However, it eventually reaches values higher than for the extensional flow.
This is consistent with the distributions of the extensions
that have been observed in experiments \cite{sbc99,stsc05a} and numerical simulations \cite{stsc05b,cpt05}.
The aforementioned tumbling events indeed entail continuous recoiling and restretching of the polymer. Therefore,
fairly large Wi are required to strech polymers appreciably,  and since the tumbling frequency increases with Wi \cite{stsc05a,tbsc05,gs06},
the distribution of the extensions becomes broader and broader as Wi grows.
A pronounced maximum at extensions comparable to $L$ only forms for Wi
as large as 200~\cite{cpt05}, and only then is $S_\rho$  expected to start decreasing.

Coming to the random case, 
$S_\rho$ displays a maximum for both the BK flow and isotropic turbulence.
At small and moderate Wi, the two curves are remarkably close despite the idealization of the BK flow.
It has indeed been shown in Ref.~\cite{mv11} that
the shape of $P(R)$ and the exponent of the power-law intermediate region 
[$P(R) \sim R^{-1-\alpha}$ for $R_0 \ll R \ll L$] are
largely insensitive to the correlation time of the flow up to correlation times of the order of $\lambda^{-1}$.
At large Wi, the behavior differs:
$S_\rho$ saturates in the BK flow, whereas it decreases in isotropic turbulence. 
The reason for this is that if the flow is turbulent and Wi is sufficiently large,
$P(R)$ displays a  power-law intermediate region together with peak near to $L$ \cite{wg10}.
The development of this sharp peak causes the reduction of $S_\rho$ at increasing $\Wi$.
In contrast, such a peak is absent in the 
BK flow, because a time-decorrelated velocity field is less effective in stretching polymers
up to their maximum length \cite{mav05} .

Figure~\ref{fig:entropy} (left panel) also shows a qualitative comparison 
with the experimental data of Sultanov {\it et al.}~\cite{ssflls21}.
This comparison requires some caveats. First of all, 
the experimental points have been translated vertically, which corresponds 
to using the extensibility parameter $b$ of the dumbbell model as 
fitting parameter \cite{psc97,lpsc97}. 
Indeed, the entropy $S_\rho$  defined from the PDF of the rescaled elongation $\rho=R/L$  
can be expressed in terms of the entropy of $P(R/R_0)$ as 
$S_\rho = S_{R/R_0} - \ln(b)/2$, where  $S_{R/R_0} = \int P(R/R_0) \ln(P(R/R_0))\, d(R/R_0)$,  
therefore a vertical translation of the entropy is equivalent to a change of $b$. 
In particular, the observation that $S_\rho^{(dumb)} \simeq S_\rho^{(exp.)} + \Delta S_\rho$ corresponds to 
fitting the experimental data with a dumbbell with equivalent extensibility
$b^{(fit)} = b^{(exp.)} [\exp( -\Delta S_\rho)]^2$. 
Thanks to this simple relation, the comparison of the entropy curves provides a useful tool for determining
the parameter $b$ of the dumbbell model which fits the experimental data.   
A precise, quantitative comparison between the experiment and the theory 
is not possible because the Weissenberg number was defined in a different way in the two cases. 
However, the analysis shows that the experimental data are qualitatively compatible with the entropy of a dumbbell in a random flow with extensibility parameter $b \approx 30^2$. 
The latter estimate is obtained from the entropy shift $\Delta S_\rho =0.33$. 
The corresponding value of the ratio $(L/R_g)^{(fit)} \approx 34.5$ 
is not far from the experimental value $(L/R_g)^{(exp.)}=47.8$.

Let us now come back to the comparison between the entropy curves in the random flows and the extensional flow. 
In both cases, the maximum of $S_\rho$ is an indication of an increased randomness 
of the polymer configuration in the transitional regime. 
However, there are some important differences in the behavior of $S_\rho$ 
observed in random flows with respect to that of the extensional flow.
First, for a comparable value of Wi the entropy is always greater in random flows. 
This is because in random flows $P(\rho)$ has a power-law intermediate region and is therefore broader.
Second, the maximum of $S_\rho$ is much wider, since in random flows the transition from the
coiled to the stretched state is much less sharp \cite{gcs05}.
Third, the maximum of $S_\rho$ is located at a value of $\Wi$ larger than $\Wicr=1/2$.
To understand this latter point, it is necessary to examine the power-law behaviour of $P(\rho)$.

As mentioned earlier, in random flows the $P(\rho)$ displays a power-law 
in the intermediate region $1/\sqrt{b}\ll \rho\ll 1$ which 
scales as $P(\rho) \sim \rho^{-1-\alpha}$,  where the exponent $\alpha$ turns from positive to negative
at $\Wicr$. Therefore, at the transition $P(\rho)\sim \rho^{-1}$. 
Given that $\alpha$ decreases monotonically with $\Wi$,
it is rather at $\Wi > \Wicr$ that $P(\rho)\sim \rho^0$ and the PDF of $\rho$ is the broadest \cite{mav05,wg10}.
Since $S_\rho$ is a measure of the randomness of $\rho$, 
it is therefore natural that in random flows $S_\rho$ reaches its maximum value at $\Wi > \Wicr$.
This fact explains the behavior of $S_\rho$.  However, it also raises the issue of
an apparent discrepancy between the critical Wi for the coil-stretch transition and the value of Wi
at which $S_\rho$ is maximum. How to reconcile these two different thresholds?

{\color{black}The time-dependent PDF $P(\bm\rho,t)$ satisfies the diffusion equation
\begin{equation}
\dfrac{\partial P}{\partial T} = - \frac{\partial}{\partial \rho_i}
\{[\kappa_{ij}(t)\rho_j-f(L\rho)\rho_i] P\}+\frac{1}{b}\, \Delta_{\bm\rho} P,
\label{eq:diffusion}
\end{equation}%
where time has been rescaled as $T=t/2\tau$ \cite{bird,o96}.
\textcolor{black}{In a statistically isotropic flow and
after the initial transient, the PDF of the rescaled extension can be assumed to depend only
on the polymer length and not on the polymer orientation
(this is true at any time if the initial PDF of $\bm\rho$ is independent of the polymer orientation).
It is therefore convenient to move to spherical coordinates (see Ref.~\cite{r89} for the transformation of the diffusion
equation under a change of variables)
and drop the derivatives
with respect to the angular variables.
This turns the relaxation and Laplacian terms into $\partial_\rho[\rho f(L\rho)P]$
and $b^{-1}\partial_\rho\rho^2\partial_\rho(P/\rho^2)$, respectively.}
The flow term can be modelled 
\textit{\`a la} Richardson, \textit{i.e.}~by describing the stretching effect 
on the polymer as a diffusion
with $\rho$-dependent eddy diffusivity \cite{sc09}. 
For a smooth random flow (recall that even in turbulent flows polymers generally lie in the dissipation
range, where the velocity field is smooth), the eddy diffusivity must be proportional to $\rho^2$ \cite{fgv01}.
In summary, moving to spherical coordinates, assuming that the solution of Eq.~\eqref{eq:diffusion} only depends on $\rho$,
and modelling the flow term via an eddy diffusivity proportional to $\rho^2$ yields
the following equation for $P(\rho,t)$:}
\begin{equation}
\dfrac{\partial P}{\partial T}=\frac{\partial}{\partial \rho}[\rho f(L\rho) P]+
\frac{\partial}{\partial \rho}\rho^2 \mathcal{K}(\rho)\frac{\partial}{\partial \rho} \frac{P}{\rho^2}
\label{eq:eddy}
\end{equation}
with $\mathcal{K}(\rho)=K \rho^2+b^{-1}$. The coefficient $K$ depends on the
the Reynolds and Weissenberg numbers in a way that is specific to the particular random flow.
However, its explicit expression is not needed for the discussion below.

Equation~\eqref{eq:eddy} can be recast as a Fokker-Planck equation
%\begin{equation}
%\dfrac{\partial P}{\partial T} = - \frac{\partial }{\partial R}[D_1(\rho)P]+\frac{\partial^2}{\partial R^2}[D_2(\rho)P]
%\label{eq:fpe}
%\end{equation}
with drift coefficient $D_1(\rho)=4K\rho-\rho f(L\rho)+2/b\rho$ and diffusion coefficient $D_2(\rho)=K\rho^2+b^{-1}$. The associated It\^o stochastic equation is
\begin{equation}
\label{eq:stochastic}
\dot{\rho}= D_1(\rho)+\sqrt{2D_2(\rho)}\, \eta(t), 
\end{equation}
where $\eta(t)$ is white noise.
Note that, for the BK flow, Eqs.~\eqref{eq:eddy} and~\eqref{eq:stochastic} hold exactly with $K=2\Wi/3$  \cite{c00,mav05}.
One important property of Eq.~\eqref{eq:stochastic} is that the amplitude of the noise depends on $\rho$.
This follows from the fact that if the flow is random, the velocity gradient 
in Eq.~\eqref{eq:dumbbell} plays the role of a multiplicative noise. 
However, to be able to use Wi as a control parameter for the coil-stretch
transition, it is
desirable to move to a representation where the amplitude
of the noise is independent of the stochastic variable, \textit{i.e.}~a stochastic equation with
additive noise only. 
This is achieved by considering a transformation of variable of the form \cite{r89}:
\begin{equation}
y\propto \int \frac{d\rho}{\sqrt{D_2(\rho)}}
= \frac{1}{\sqrt{K}}\,\ln[K\rho+\sqrt{K(K\rho^2+b^{-1})}] + \text{const.}
\label{eq:y-exact}
\end{equation}
Around the coil-stretch transition, the coefficient $K$ is $O(1)$. For $\rho\gg 1/\sqrt{b}$ Eq.~\eqref{eq:y-exact} thus gives
\begin{equation}
y\sim\ln\rho.
\label{eq:y-ln}
\end{equation}
Now note that the PDF of $y$ is related to that of $\rho$ via the relationship $P(y)\propto \rho\,P(\rho)$. Therefore, according to
the theory of Balkovsky \textit{et al.} \cite{bfl00}, at $\Wi=\Wicr$ the power-law region of $P(y)$ 
is flat and the entropy of $y$,
\begin{equation}
\label{eq:entropy-y}
S_y=-\int P(y)\, \ln[P(y)]\, dy,
\end{equation}
is expected to reach its maximum value. This suggests that, for random flows, it may be more appropriate to characterize the 
coil-stretch transition by measuring the entropy of $y$ rather than that of $\rho$.
(The logarithm of the polymer extension has also been used in other contexts, 
for instance to improve accuracy in numerical simulations of constitutive models of polymer solutions \cite{vc03,fk04} or to develop a geometric decomposition of the conformation tensor $\mathbf{C}=\langle\bm \rho\otimes\bm\rho\rangle_\xi$ that guarantees the positive definiteness of both its mean and fluctuating components \cite{hmzg18}.)

%\begin{figure}[!t]
%\centering
%\includegraphics[width=0.5\columnwidth]{plot_entropy_log}
%\caption{Entropy of $y=\ln\rho$ for the BK and turbulent flows and for
%the same parameters as in Fig.~\ref{fig:entropy}.}
%\label{fig:log}
%\end{figure}

Figure~\ref{fig:entropy} (right panel) shows $S_y$ vs Wi for the BK flow and  isotropic turbulence. 
The experimental data have not been included because
calculating $P(y)$ from $P(\rho)$ would require a higher resolution of the small extensions
than that available in the experiment [recall that $P(y)\sim \rho\,P(\rho)$].
As expected, $S_y$ is maximum at $\Wi=\Wicr$,
which confirms that in random flows $S_y$ provides a convenient characterization of the coil-stretch transition.
The differences between the BK flow and isotropic turbulence that have been discussed earlier obviously also
manifest themselves in the behaviour of $S_y$. 

\textcolor{black}{Previous studies have investigated the correlation between the polymer extension and the local flow topology
 \cite{terrapon,ps07,wg10}.
In a three-dimensional turbulent flow, 
the sign of the second invariant of the velocity gradient, $Q=-\operatorname{tr}(\nabla\bm u)^2/2$, discriminates
between the regions of the flow that are dominated by strain ($Q<0$) and those that are dominated by vorticity ($Q>0$) \cite{bmc96}.
 In order to determine the dependence of entropy on the local flow topology,
 we consider the conditional probabilities $P^-(y)=P(y |Q<0)$ and $P^+(y)=P(y |Q>0)$ and the associated
 entropies $S^-_y$ and $S^+_y$, respectively. These are shown in the inset of Fig.~\ref{fig:entropy} (right panel) as a function
 of the Weissenberg number. $S_y^-$ is obviously greater than $S_y^+$ at large Wi, but the difference between the two entropies is not big. 
This is consistent with the fact that the extension of a polymer depends on its stretching history 
and not only on the instantaneous velocity gradient.
 }

\section{Summary and conclusions}

In a non-uniform flow, polymers can be highly deformed by the local velocity gradients.
However, the statistics of the deformation and the way it varies with Wi depend
very sensitively on the properties of the flow. In particular, substantial
differences are observed between laminar and random velocity fields.
An entropic characterization of the coil-stretch transition 
{\color{black}was proposed
by Latinwo {\it et. al.}~\cite{latinwo2014nonequilibrium} for an extensional flow.
This characterization has been recently extended to random flows by Sultanov \textit{et al.}~\cite{ssflls21}}.  
We have further developped this approach by examining 
a set of flows that have been regarded as benchmarks for
the study of polymer stretching, in both the laminar and the random case.
This study confirms that the dependence of entropy on Wi provides
a useful characterization of the change in the statistics of polymer extension that occurs near 
the coil-stretch transition. Moreover, it allows a quantitative comparison between flows with different
stretching properties. 
This characterization is particularly relevant to practical situations where limited
statistics is available. Entropy is indeed less sensitive to statistical fluctuations than quantities, 
such as the slope of $P(\rho)$ or the correlation time of $\rho(t)$, which have been used previously
to describe the coil-stretch transition.

\acknowledgments
The authors are grateful to Samriddhi Sankar Ray for useful discussions and for providing access to his database of turbulent Lagrangian trajectories.
V.S. acknowledges financial support from the Israel Science Foundation (ISF grant \#784/19).
D.V. acknowledges the support of the French government through the UCAJEDI Investments in the Future project managed by the National Research Agency 
(ANR) under reference number ANR-15-IDEX-01 and thanks
the OPAL infrastructure and the Center for High-Performance Computing of Universit\'e Côte d'Azur for computational resources.
D.V. also acknowledges support from the International Centre for Theoretical Sciences (ICTS-TIFR), Bangalore, India.

{\color{black}
\appendix

\section{}
\label{appendix}

Consider a chain with $N$ beads and $N-1$ springs. If the position of the $i$-th bead is denoted as $\mathbf{x}_i$, 
the connectors $\mathbf{Q}_{i}=\mathbf{x}_{i+1}-\mathbf{x}_{i}$ ($i=1,\dots,N-1$) satisfy
\begin{equation}
\begin{array}{lcl}
\dot{\mathbf{Q}}_1 &=&  \bm\kappa(t)\cdot\mathbf{Q}_1 - \dfrac{1}{4\tau}(2 f_1 \mathbf{Q}_1 - f_{2}\mathbf{Q}_{2})
+\dfrac{Q_{\rm eq}^c}{\sqrt{6\tau}} \, [\bm\xi_{2}(t)-\bm\xi_1(t)],
\\[4mm]
\dot{\mathbf{Q}}_i &=& \bm\kappa(t)\cdot\mathbf{Q}_i - \dfrac{1}{4\tau}(2 f_i \mathbf{Q}_i - f_{i+1}\mathbf{Q}_{i+1} - f_{i-1}\mathbf{Q}_{i-1})
+\dfrac{Q_{\rm eq}^c}{\sqrt{6\tau}} \, [\bm\xi_{i+1}(t)-\bm\xi_i(t)], \qquad (i=2,\dots,N-2)
\\[4mm]
\dot{\mathbf{Q}}_{N-1} &=&  \bm\kappa(t)\cdot\mathbf{Q}_{N-1} - \dfrac{1}{4\tau}(2 f_{N-1} \mathbf{Q}_{N-1} - f_{N-2}\mathbf{Q}_{N-2})
+\dfrac{Q_{\rm eq}^c}{\sqrt{6\tau}} \, [\bm\xi_{N}(t)-\bm\xi_{N-1}(t)],
\end{array}
\label{eq:chain}
\end{equation}
where $\tau^c$ and $Q_{\mathrm{eq}}^c$ are the relaxation time and equilibrium  length of the springs, respectively,
and $\bm\xi_i(t)$ are independent three-dimensional white noises.
The coefficients
\begin{equation}
f_i = \frac{1}{1-(Q_i/Q_{\mathrm{max}}^c)^2}
\end{equation}
describe the nonlinear elasticity of the springs and fix the maximum length of each of them to $Q_{\mathrm{max}}^c$.
Therefore the maximum length of the chain is $L^c = Q_{\mathrm{max}}^c (N-1)$.

Equations \eqref{eq:chain} have been solved by means of the Euler-Maruyama method supplemented with
\"Ottinger's rejection algorithm, which rejects those time steps for which there exists at least
one index $i$ such that
$\vert\mathbf{Q}_i\vert>Q_\mathrm{max}^c(1-\sqrt{dt/10\tau})^{1/2}$ \cite{o96}.
We have checked that the fraction of rejected time steps was
negligible for all Wi and for all flows considered here.

The end-to-end separation vector of the chain is $\mathbf{R}=\sum_{i=1}^{N-1}\mathbf{Q}_i$.
In order to compare the results for a dumbbell ($N=2$) with those for a multi-bead chain ($N>2$), we have
used the mapping proposed in Ref.~\cite{jc07}. This mapping assumes that 
the statistics of the end-to-end separation of a $N$-bead chain is equivalent to that of a dumbbell with parameters:
\begin{equation}
\tau = \frac{N(N+1)\tau^c}{6}, \qquad R_{\mathrm{eq}} = Q_{\mathrm{eq}}^c, \qquad L = Q_{\mathrm{max}}^c \sqrt{N-1} = L^c/\sqrt{N-1}.
\end{equation}
In Fig.~\ref{fig:chain}, we plot the entropy of the rescaled end-to-end separation $\rho=R/L^c$ as a function of Wi for the uniaxial extensional
flow, the linear shear flow, and isotropic turbulence. The curves for the dumbbell model shown in Fig.~\ref{fig:entropy}(a) are compared with those for
an equivalent chain with $N=10$ beads. Only small quantitative differences are observed 
between the results for $N=2$ and $N=10$.
\begin{figure}
\setlength{\unitlength}{\textwidth}
\includegraphics[width=0.33\textwidth]{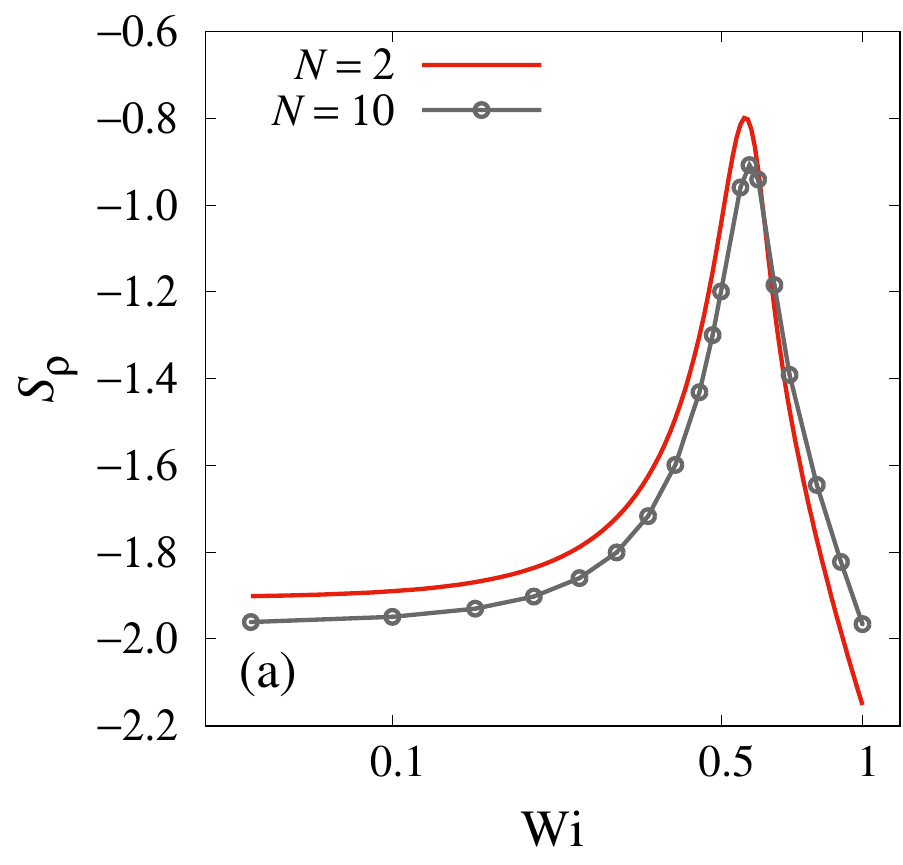}\hfill%
\includegraphics[width=0.33\textwidth]{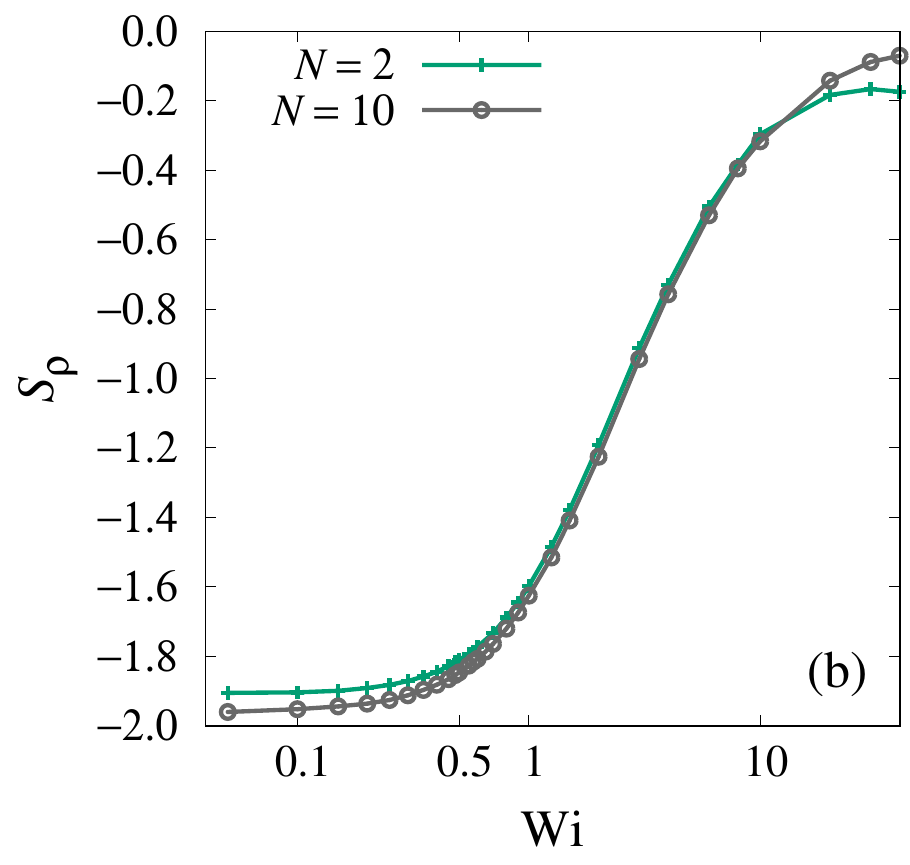}\hfill%
\includegraphics[width=0.335\textwidth]{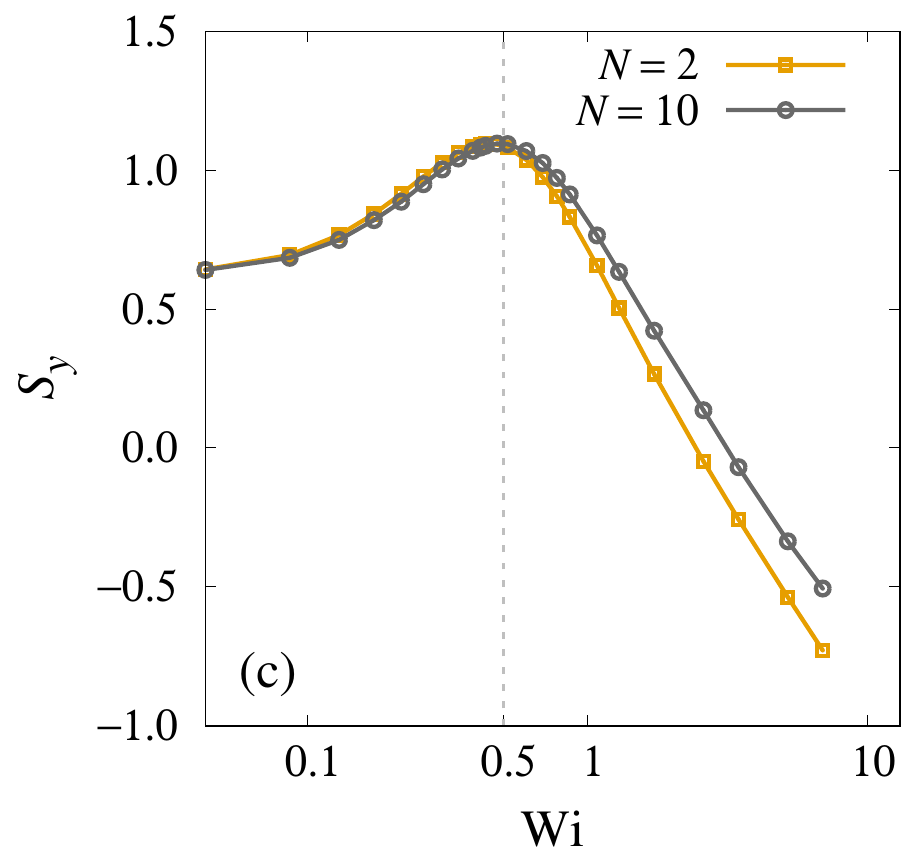}\hfill%
\caption{\color{black} Entropy of the fractional end-to-end separation as a function of Wi for a dumbbell and a chain with $N=10$ in (a) a 
uniaxial extensional flow, (b) a linear shear flow, and (c) isotropic turbulence.}
\label{fig:chain}
\end{figure}

}

%\nocite{*}
\bibliography{polymers}% Produces the bibliography via BibTeX.

\begin{thebibliography}{72}%
\makeatletter
\providecommand \@ifxundefined [1]{%
 \@ifx{#1\undefined}
}%
\providecommand \@ifnum [1]{%
 \ifnum #1\expandafter \@firstoftwo
 \else \expandafter \@secondoftwo
 \fi
}%
\providecommand \@ifx [1]{%
 \ifx #1\expandafter \@firstoftwo
 \else \expandafter \@secondoftwo
 \fi
}%
\providecommand \natexlab [1]{#1}%
\providecommand \enquote  [1]{``#1''}%
\providecommand \bibnamefont  [1]{#1}%
\providecommand \bibfnamefont [1]{#1}%
\providecommand \citenamefont [1]{#1}%
\providecommand \href@noop [0]{\@secondoftwo}%
\providecommand \href [0]{\begingroup \@sanitize@url \@href}%
\providecommand \@href[1]{\@@startlink{#1}\@@href}%
\providecommand \@@href[1]{\endgroup#1\@@endlink}%
\providecommand \@sanitize@url [0]{\catcode `\\12\catcode `\$12\catcode
  `\&12\catcode `\#12\catcode `\^12\catcode `\_12\catcode `\%12\relax}%
\providecommand \@@startlink[1]{}%
\providecommand \@@endlink[0]{}%
\providecommand \url  [0]{\begingroup\@sanitize@url \@url }%
\providecommand \@url [1]{\endgroup\@href {#1}{\urlprefix }}%
\providecommand \urlprefix  [0]{URL }%
\providecommand \Eprint [0]{\href }%
\providecommand \doibase [0]{https://doi.org/}%
\providecommand \selectlanguage [0]{\@gobble}%
\providecommand \bibinfo  [0]{\@secondoftwo}%
\providecommand \bibfield  [0]{\@secondoftwo}%
\providecommand \translation [1]{[#1]}%
\providecommand \BibitemOpen [0]{}%
\providecommand \bibitemStop [0]{}%
\providecommand \bibitemNoStop [0]{.\EOS\space}%
\providecommand \EOS [0]{\spacefactor3000\relax}%
\providecommand \BibitemShut  [1]{\csname bibitem#1\endcsname}%
\let\auto@bib@innerbib\@empty
%</preamble>
\bibitem [{\citenamefont {{P.-G. de Gennes}}(1974)}]{deGennes}%
  \BibitemOpen
  \bibfield  {author} {\bibinfo {author} {\bibnamefont {{P.-G. de Gennes}}},\
  }\bibfield  {title} {\bibinfo {title} {Coil‐stretch transition of dilute
  flexible polymers under ultrahigh velocity gradients},\ }\href@noop {}
  {\bibfield  {journal} {\bibinfo  {journal} {J. Chem. Phys.}\ }\textbf
  {\bibinfo {volume} {60}},\ \bibinfo {pages} {5030} (\bibinfo {year}
  {1974})}\BibitemShut {NoStop}%
\bibitem [{\citenamefont {Perkins}\ \emph {et~al.}(1997)\citenamefont
  {Perkins}, \citenamefont {Smith},\ and\ \citenamefont {Chu}}]{psc97}%
  \BibitemOpen
  \bibfield  {author} {\bibinfo {author} {\bibfnamefont {T.}~\bibnamefont
  {Perkins}}, \bibinfo {author} {\bibfnamefont {D.~E.}\ \bibnamefont {Smith}},\
  and\ \bibinfo {author} {\bibfnamefont {S.}~\bibnamefont {Chu}},\ }\bibfield
  {title} {\bibinfo {title} {Single polymer dynamics in an elongational flow},\
  }\href@noop {} {\bibfield  {journal} {\bibinfo  {journal} {Science}\ }\textbf
  {\bibinfo {volume} {276}},\ \bibinfo {pages} {2016} (\bibinfo {year}
  {1997})}\BibitemShut {NoStop}%
\bibitem [{\citenamefont {Schroeder}\ \emph {et~al.}(2003)\citenamefont
  {Schroeder}, \citenamefont {Babcock}, \citenamefont {Shaqfeh},\ and\
  \citenamefont {Chu}}]{sbsc03}%
  \BibitemOpen
  \bibfield  {author} {\bibinfo {author} {\bibfnamefont {C.~M.}\ \bibnamefont
  {Schroeder}}, \bibinfo {author} {\bibfnamefont {H.~P.}\ \bibnamefont
  {Babcock}}, \bibinfo {author} {\bibfnamefont {E.~S.~G.}\ \bibnamefont
  {Shaqfeh}},\ and\ \bibinfo {author} {\bibfnamefont {S.}~\bibnamefont {Chu}},\
  }\bibfield  {title} {\bibinfo {title} {Observation of polymer conformation
  hysteresis in extensional flow},\ }\href@noop {} {\bibfield  {journal}
  {\bibinfo  {journal} {Science}\ }\textbf {\bibinfo {volume} {301}},\ \bibinfo
  {pages} {1515} (\bibinfo {year} {2003})}\BibitemShut {NoStop}%
\bibitem [{\citenamefont {Gerashchenko}\ \emph {et~al.}(2005)\citenamefont
  {Gerashchenko}, \citenamefont {Chevallard},\ and\ \citenamefont
  {Steinberg}}]{gcs05}%
  \BibitemOpen
  \bibfield  {author} {\bibinfo {author} {\bibfnamefont {S.}~\bibnamefont
  {Gerashchenko}}, \bibinfo {author} {\bibfnamefont {C.}~\bibnamefont
  {Chevallard}},\ and\ \bibinfo {author} {\bibfnamefont {V.}~\bibnamefont
  {Steinberg}},\ }\bibfield  {title} {\bibinfo {title} {Single-polymer
  dynamics: Coil-stretch transition in a random flow},\ }\href@noop {}
  {\bibfield  {journal} {\bibinfo  {journal} {Europhys. Lett.}\ }\textbf
  {\bibinfo {volume} {71}},\ \bibinfo {pages} {221} (\bibinfo {year}
  {2005})}\BibitemShut {NoStop}%
\bibitem [{\citenamefont {Liu}\ and\ \citenamefont {Steinberg}(2010)}]{ls10}%
  \BibitemOpen
  \bibfield  {author} {\bibinfo {author} {\bibfnamefont {Y.}~\bibnamefont
  {Liu}}\ and\ \bibinfo {author} {\bibfnamefont {V.}~\bibnamefont
  {Steinberg}},\ }\bibfield  {title} {\bibinfo {title} {Stretching of polymer
  in a random flow: Effect of a shear rate},\ }\href@noop {} {\bibfield
  {journal} {\bibinfo  {journal} {Europhys. Lett.}\ }\textbf {\bibinfo {volume}
  {90}},\ \bibinfo {pages} {44005} (\bibinfo {year} {2010})}\BibitemShut
  {NoStop}%
\bibitem [{\citenamefont {Liu}\ and\ \citenamefont {Steinberg}(2014)}]{ls14}%
  \BibitemOpen
  \bibfield  {author} {\bibinfo {author} {\bibfnamefont {Y.}~\bibnamefont
  {Liu}}\ and\ \bibinfo {author} {\bibfnamefont {V.}~\bibnamefont
  {Steinberg}},\ }\bibfield  {title} {\bibinfo {title} {Single polymer dynamics
  in a random flow},\ }\href@noop {} {\bibfield  {journal} {\bibinfo  {journal}
  {Macromol. Symp.}\ }\textbf {\bibinfo {volume} {337}},\ \bibinfo {pages} {34}
  (\bibinfo {year} {2014})}\BibitemShut {NoStop}%
\bibitem [{\citenamefont {Tang}\ \emph {et~al.}(2010)\citenamefont {Tang},
  \citenamefont {Trahan},\ and\ \citenamefont {Doyle}}]{ttd10}%
  \BibitemOpen
  \bibfield  {author} {\bibinfo {author} {\bibfnamefont {J.}~\bibnamefont
  {Tang}}, \bibinfo {author} {\bibfnamefont {D.~W.}\ \bibnamefont {Trahan}},\
  and\ \bibinfo {author} {\bibfnamefont {P.~S.}\ \bibnamefont {Doyle}},\
  }\bibfield  {title} {\bibinfo {title} {Coil--stretch transition of {DNA}
  molecules in slitlike confinement},\ }\href@noop {} {\bibfield  {journal}
  {\bibinfo  {journal} {Macromolecules}\ }\textbf {\bibinfo {volume} {43}},\
  \bibinfo {pages} {3081} (\bibinfo {year} {2010})}\BibitemShut {NoStop}%
\bibitem [{\citenamefont {Somani}\ \emph {et~al.}(2010)\citenamefont {Somani},
  \citenamefont {Shaqfeh},\ and\ \citenamefont {Prakash}}]{ssp10}%
  \BibitemOpen
  \bibfield  {author} {\bibinfo {author} {\bibfnamefont {S.}~\bibnamefont
  {Somani}}, \bibinfo {author} {\bibfnamefont {E.~S.~G.}\ \bibnamefont
  {Shaqfeh}},\ and\ \bibinfo {author} {\bibfnamefont {J.~R.}\ \bibnamefont
  {Prakash}},\ }\bibfield  {title} {\bibinfo {title} {Effect of solvent quality
  on the coil--stretch transition},\ }\href@noop {} {\bibfield  {journal}
  {\bibinfo  {journal} {Macromolecules}\ }\textbf {\bibinfo {volume} {43}},\
  \bibinfo {pages} {10679} (\bibinfo {year} {2010})}\BibitemShut {NoStop}%
\bibitem [{\citenamefont {Radhakrishnan}\ and\ \citenamefont
  {Underhill}(2012)}]{ru12}%
  \BibitemOpen
  \bibfield  {author} {\bibinfo {author} {\bibfnamefont {R.}~\bibnamefont
  {Radhakrishnan}}\ and\ \bibinfo {author} {\bibfnamefont {P.~T.}\ \bibnamefont
  {Underhill}},\ }\bibfield  {title} {\bibinfo {title} {Models of flexible
  polymers in good solvents: relaxation and coil-stretch transition},\
  }\href@noop {} {\bibfield  {journal} {\bibinfo  {journal} {Soft Matter}\
  }\textbf {\bibinfo {volume} {8}},\ \bibinfo {pages} {6991} (\bibinfo {year}
  {2012})}\BibitemShut {NoStop}%
\bibitem [{\citenamefont {Radhakrishnan}\ and\ \citenamefont
  {Underhill}(2013)}]{ru13}%
  \BibitemOpen
  \bibfield  {author} {\bibinfo {author} {\bibfnamefont {R.}~\bibnamefont
  {Radhakrishnan}}\ and\ \bibinfo {author} {\bibfnamefont {P.~T.}\ \bibnamefont
  {Underhill}},\ }\bibfield  {title} {\bibinfo {title} {Impact of solvent
  quality on the hysteresis in the coil-stretch transition of flexible polymers
  in good solvents},\ }\href@noop {} {\bibfield  {journal} {\bibinfo  {journal}
  {Macromolecules}\ }\textbf {\bibinfo {volume} {46}},\ \bibinfo {pages} {548}
  (\bibinfo {year} {2013})}\BibitemShut {NoStop}%
\bibitem [{\citenamefont {Prabhakar}\ \emph {et~al.}(2017)\citenamefont
  {Prabhakar}, \citenamefont {Sasmal}, \citenamefont {Nguyen}, \citenamefont
  {Sridhar},\ and\ \citenamefont {Prakash}}]{prabhakar}%
  \BibitemOpen
  \bibfield  {author} {\bibinfo {author} {\bibfnamefont {R.}~\bibnamefont
  {Prabhakar}}, \bibinfo {author} {\bibfnamefont {C.}~\bibnamefont {Sasmal}},
  \bibinfo {author} {\bibfnamefont {D.~A.}\ \bibnamefont {Nguyen}}, \bibinfo
  {author} {\bibfnamefont {T.}~\bibnamefont {Sridhar}},\ and\ \bibinfo {author}
  {\bibfnamefont {J.~R.}\ \bibnamefont {Prakash}},\ }\bibfield  {title}
  {\bibinfo {title} {Effect of stretching-induced changes in hydrodynamic
  screening on coil-stretch hysteresis of unentangled polymer solutions},\
  }\href@noop {} {\bibfield  {journal} {\bibinfo  {journal} {Phys. Rev.
  Fluids}\ }\textbf {\bibinfo {volume} {2}},\ \bibinfo {pages} {011301(R)}
  (\bibinfo {year} {2017})}\BibitemShut {NoStop}%
\bibitem [{\citenamefont {Soh}\ \emph {et~al.}(2018)\citenamefont {Soh},
  \citenamefont {Narsimhan}, \citenamefont {Klotz},\ and\ \citenamefont
  {Doyle}}]{snkd18}%
  \BibitemOpen
  \bibfield  {author} {\bibinfo {author} {\bibfnamefont {B.~W.}\ \bibnamefont
  {Soh}}, \bibinfo {author} {\bibfnamefont {V.}~\bibnamefont {Narsimhan}},
  \bibinfo {author} {\bibfnamefont {A.~R.}\ \bibnamefont {Klotz}},\ and\
  \bibinfo {author} {\bibfnamefont {P.~S.}\ \bibnamefont {Doyle}},\ }\bibfield
  {title} {\bibinfo {title} {Knots modify the coil–stretch transition in
  linear {DNA} polymers},\ }\href@noop {} {\bibfield  {journal} {\bibinfo
  {journal} {Soft Matter}\ }\textbf {\bibinfo {volume} {14}},\ \bibinfo {pages}
  {1689} (\bibinfo {year} {2018})}\BibitemShut {NoStop}%
\bibitem [{\citenamefont {Cifre}\ \emph {et~al.}(2005)\citenamefont {Cifre},
  \citenamefont {Pamies}, \citenamefont {Martinez},\ and\ \citenamefont {de~la
  Torre}}]{delatorre2005}%
  \BibitemOpen
  \bibfield  {author} {\bibinfo {author} {\bibfnamefont {J.~G.~H.}\
  \bibnamefont {Cifre}}, \bibinfo {author} {\bibfnamefont {R.}~\bibnamefont
  {Pamies}}, \bibinfo {author} {\bibfnamefont {M.~C.~L.}\ \bibnamefont
  {Martinez}},\ and\ \bibinfo {author} {\bibfnamefont {J.~G.}\ \bibnamefont
  {de~la Torre}},\ }\bibfield  {title} {\bibinfo {title} {Steady-state behavior
  of ring polymers in dilute flowing solutions via {Brownian Dynamics}},\
  }\href@noop {} {\bibfield  {journal} {\bibinfo  {journal} {Polymer}\ }\textbf
  {\bibinfo {volume} {46}},\ \bibinfo {pages} {267} (\bibinfo {year}
  {2005})}\BibitemShut {NoStop}%
\bibitem [{\citenamefont {Li}\ \emph {et~al.}(2015)\citenamefont {Li},
  \citenamefont {Hsiao}, \citenamefont {Brockman}, \citenamefont {Yates},
  \citenamefont {Robertson-Anderson}, \citenamefont {Kornfield}, \citenamefont
  {Francisco}, \citenamefont {Schroeder},\ and\ \citenamefont
  {McKenna}}]{rings}%
  \BibitemOpen
  \bibfield  {author} {\bibinfo {author} {\bibfnamefont {Y.}~\bibnamefont
  {Li}}, \bibinfo {author} {\bibfnamefont {K.-W.}\ \bibnamefont {Hsiao}},
  \bibinfo {author} {\bibfnamefont {C.~A.}\ \bibnamefont {Brockman}}, \bibinfo
  {author} {\bibfnamefont {D.~Y.}\ \bibnamefont {Yates}}, \bibinfo {author}
  {\bibfnamefont {R.~M.}\ \bibnamefont {Robertson-Anderson}}, \bibinfo {author}
  {\bibfnamefont {J.~A.}\ \bibnamefont {Kornfield}}, \bibinfo {author}
  {\bibfnamefont {M.~J.~S.}\ \bibnamefont {Francisco}}, \bibinfo {author}
  {\bibfnamefont {C.~M.}\ \bibnamefont {Schroeder}},\ and\ \bibinfo {author}
  {\bibfnamefont {G.~B.}\ \bibnamefont {McKenna}},\ }\bibfield  {title}
  {\bibinfo {title} {When ends meet: Circular {DNA} stretches differently in
  elongational flows},\ }\href@noop {} {\bibfield  {journal} {\bibinfo
  {journal} {Macromolecules}\ }\textbf {\bibinfo {volume} {48}},\ \bibinfo
  {pages} {5997} (\bibinfo {year} {2015})}\BibitemShut {NoStop}%
\bibitem [{\citenamefont {Hsiao}\ \emph {et~al.}(2016)\citenamefont {Hsiao},
  \citenamefont {Schroeder},\ and\ \citenamefont {Sing}}]{hss16}%
  \BibitemOpen
  \bibfield  {author} {\bibinfo {author} {\bibfnamefont {K.-W.}\ \bibnamefont
  {Hsiao}}, \bibinfo {author} {\bibfnamefont {C.~M.}\ \bibnamefont
  {Schroeder}},\ and\ \bibinfo {author} {\bibfnamefont {C.~E.}\ \bibnamefont
  {Sing}},\ }\bibfield  {title} {\bibinfo {title} {Ring polymer dynamics are
  governed by a coupling between architecture and hydrodynamic interactions},\
  }\href@noop {} {\bibfield  {journal} {\bibinfo  {journal} {Macromolecules}\
  }\textbf {\bibinfo {volume} {49}},\ \bibinfo {pages} {1961} (\bibinfo {year}
  {2016})}\BibitemShut {NoStop}%
\bibitem [{\citenamefont {{Nafar Sefiddashti}}\ \emph
  {et~al.}(2018{\natexlab{a}})\citenamefont {{Nafar Sefiddashti}},
  \citenamefont {Edwards},\ and\ \citenamefont {Khomami}}]{sek18b}%
  \BibitemOpen
  \bibfield  {author} {\bibinfo {author} {\bibfnamefont {M.~H.}\ \bibnamefont
  {{Nafar Sefiddashti}}}, \bibinfo {author} {\bibfnamefont {B.~J.}\
  \bibnamefont {Edwards}},\ and\ \bibinfo {author} {\bibfnamefont
  {B.}~\bibnamefont {Khomami}},\ }\bibfield  {title} {\bibinfo {title}
  {Communication: A coil-stretch transition in planar elongational flow of an
  entangled polymeric melt},\ }\href@noop {} {\bibfield  {journal} {\bibinfo
  {journal} {J. Chem. Phys.}\ }\textbf {\bibinfo {volume} {148}},\ \bibinfo
  {pages} {141103} (\bibinfo {year} {2018}{\natexlab{a}})}\BibitemShut
  {NoStop}%
\bibitem [{\citenamefont {{Nafar Sefiddashti}}\ \emph
  {et~al.}(2018{\natexlab{b}})\citenamefont {{Nafar Sefiddashti}},
  \citenamefont {Edwards},\ and\ \citenamefont {Khomami}}]{sek18}%
  \BibitemOpen
  \bibfield  {author} {\bibinfo {author} {\bibfnamefont {M.~H.}\ \bibnamefont
  {{Nafar Sefiddashti}}}, \bibinfo {author} {\bibfnamefont {B.~J.}\
  \bibnamefont {Edwards}},\ and\ \bibinfo {author} {\bibfnamefont
  {B.}~\bibnamefont {Khomami}},\ }\bibfield  {title} {\bibinfo {title}
  {Configurational microphase separation in elongational flow of an entangled
  polymer liquid},\ }\href@noop {} {\bibfield  {journal} {\bibinfo  {journal}
  {Phys. Rev. Lett.}\ }\textbf {\bibinfo {volume} {121}},\ \bibinfo {pages}
  {247802} (\bibinfo {year} {2018}{\natexlab{b}})}\BibitemShut {NoStop}%
\bibitem [{\citenamefont {Kirk}\ \emph {et~al.}(2018)\citenamefont {Kirk},
  \citenamefont {Kr\"oger},\ and\ \citenamefont {Ilg}}]{kki18}%
  \BibitemOpen
  \bibfield  {author} {\bibinfo {author} {\bibfnamefont {J.}~\bibnamefont
  {Kirk}}, \bibinfo {author} {\bibfnamefont {M.}~\bibnamefont {Kr\"oger}},\
  and\ \bibinfo {author} {\bibfnamefont {P.}~\bibnamefont {Ilg}},\ }\bibfield
  {title} {\bibinfo {title} {Surface disentanglement and slip in a polymer
  melt: A molecular dynamics study},\ }\href@noop {} {\bibfield  {journal}
  {\bibinfo  {journal} {Macromolecules}\ }\textbf {\bibinfo {volume} {51}},\
  \bibinfo {pages} {8996} (\bibinfo {year} {2018})}\BibitemShut {NoStop}%
\bibitem [{\citenamefont {Yu}\ and\ \citenamefont {Graham}(2021)}]{yg21}%
  \BibitemOpen
  \bibfield  {author} {\bibinfo {author} {\bibfnamefont {Y.~J.}\ \bibnamefont
  {Yu}}\ and\ \bibinfo {author} {\bibfnamefont {M.~D.}\ \bibnamefont
  {Graham}},\ }\bibfield  {title} {\bibinfo {title} {Coil-stretch-like
  transition of elastic sheets in extensional flows},\ }\href@noop {}
  {\bibfield  {journal} {\bibinfo  {journal} {Soft Matter}\ }\textbf {\bibinfo
  {volume} {17}},\ \bibinfo {pages} {543} (\bibinfo {year} {2021})}\BibitemShut
  {NoStop}%
\bibitem [{\citenamefont {Graham}(2014)}]{g14}%
  \BibitemOpen
  \bibfield  {author} {\bibinfo {author} {\bibfnamefont {M.~D.}\ \bibnamefont
  {Graham}},\ }\bibfield  {title} {\bibinfo {title} {Drag reduction and the
  dynamics of turbulence in simple and complex fluids},\ }\href@noop {}
  {\bibfield  {journal} {\bibinfo  {journal} {Phys. Fluids}\ }\textbf {\bibinfo
  {volume} {26}},\ \bibinfo {pages} {101301} (\bibinfo {year}
  {2014})}\BibitemShut {NoStop}%
\bibitem [{\citenamefont {Benzi}\ and\ \citenamefont {Ching}(2018)}]{bc18}%
  \BibitemOpen
  \bibfield  {author} {\bibinfo {author} {\bibfnamefont {R.}~\bibnamefont
  {Benzi}}\ and\ \bibinfo {author} {\bibfnamefont {E.~S.~C.}\ \bibnamefont
  {Ching}},\ }\bibfield  {title} {\bibinfo {title} {Polymers in fluid flows},\
  }\href@noop {} {\bibfield  {journal} {\bibinfo  {journal} {Annu. Rev.
  Condens. Matter Phys.}\ }\textbf {\bibinfo {volume} {9}},\ \bibinfo {pages}
  {163} (\bibinfo {year} {2018})}\BibitemShut {NoStop}%
\bibitem [{\citenamefont {Xi}(2019)}]{x19}%
  \BibitemOpen
  \bibfield  {author} {\bibinfo {author} {\bibfnamefont {L.}~\bibnamefont
  {Xi}},\ }\bibfield  {title} {\bibinfo {title} {Turbulent drag reduction by
  polymer additives: Fundamentals and recent advances},\ }\href@noop {}
  {\bibfield  {journal} {\bibinfo  {journal} {Phys. Fluids}\ }\textbf {\bibinfo
  {volume} {31}},\ \bibinfo {pages} {121302} (\bibinfo {year}
  {2019})}\BibitemShut {NoStop}%
\bibitem [{\citenamefont {Steinberg}(2021)}]{steinberg21}%
  \BibitemOpen
  \bibfield  {author} {\bibinfo {author} {\bibfnamefont {V.}~\bibnamefont
  {Steinberg}},\ }\bibfield  {title} {\bibinfo {title} {Elastic turbulence: An
  experimental view on inertialess random flow},\ }\href@noop {} {\bibfield
  {journal} {\bibinfo  {journal} {Annu. Rev. Fluid Mech.}\ }\textbf {\bibinfo
  {volume} {53}},\ \bibinfo {pages} {27} (\bibinfo {year} {2021})}\BibitemShut
  {NoStop}%
\bibitem [{\citenamefont {Steinberg}(2022)}]{steinberg22}%
  \BibitemOpen
  \bibfield  {author} {\bibinfo {author} {\bibfnamefont {V.}~\bibnamefont
  {Steinberg}},\ }\bibfield  {title} {\bibinfo {title} {New direction and
  perspectives in elastic instability and turbulence in various viscoelastic
  flow geometries without inertia},\ }\href@noop {} {\bibfield  {journal}
  {\bibinfo  {journal} {Low Temp. Phys.}\ }\textbf {\bibinfo {volume} {48}},\
  \bibinfo {pages} {492} (\bibinfo {year} {2022})}\BibitemShut {NoStop}%
\bibitem [{\citenamefont {Datta}\ \emph {et~al.}(2022)\citenamefont {Datta},
  \citenamefont {Ardekani}, \citenamefont {Arratia}, \citenamefont {Beris},
  \citenamefont {Bischofberger},\ and\ \citenamefont {{McKinley {\it et
  al.}}}}]{datta22}%
  \BibitemOpen
  \bibfield  {author} {\bibinfo {author} {\bibfnamefont {S.~S.}\ \bibnamefont
  {Datta}}, \bibinfo {author} {\bibfnamefont {A.~M.}\ \bibnamefont {Ardekani}},
  \bibinfo {author} {\bibfnamefont {P.~E.}\ \bibnamefont {Arratia}}, \bibinfo
  {author} {\bibfnamefont {A.~N.}\ \bibnamefont {Beris}}, \bibinfo {author}
  {\bibfnamefont {I.}~\bibnamefont {Bischofberger}},\ and\ \bibinfo {author}
  {\bibfnamefont {G.~H.}\ \bibnamefont {{McKinley {\it et al.}}}},\ }\bibfield
  {title} {\bibinfo {title} {Perspectives on viscoelastic flow instabilities
  and elastic turbulence},\ }\href@noop {} {\bibfield  {journal} {\bibinfo
  {journal} {Phys. Rev. Fluids}\ }\textbf {\bibinfo {volume} {7}},\ \bibinfo
  {pages} {080701} (\bibinfo {year} {2022})}\BibitemShut {NoStop}%
\bibitem [{\citenamefont {Celani}\ \emph {et~al.}(2006)\citenamefont {Celani},
  \citenamefont {Puliafito},\ and\ \citenamefont {Vincenzi}}]{cpv06}%
  \BibitemOpen
  \bibfield  {author} {\bibinfo {author} {\bibfnamefont {A.}~\bibnamefont
  {Celani}}, \bibinfo {author} {\bibfnamefont {A.}~\bibnamefont {Puliafito}},\
  and\ \bibinfo {author} {\bibfnamefont {D.}~\bibnamefont {Vincenzi}},\
  }\bibfield  {title} {\bibinfo {title} {Dynamical slowdown of polymers in
  laminar and random flows},\ }\href@noop {} {\bibfield  {journal} {\bibinfo
  {journal} {Phys. Rev. Lett.}\ }\textbf {\bibinfo {volume} {97}},\ \bibinfo
  {pages} {118301} (\bibinfo {year} {2006})}\BibitemShut {NoStop}%
\bibitem [{\citenamefont {Gerashchenko}\ and\ \citenamefont
  {Steinberg}(2008)}]{gs08}%
  \BibitemOpen
  \bibfield  {author} {\bibinfo {author} {\bibfnamefont {S.}~\bibnamefont
  {Gerashchenko}}\ and\ \bibinfo {author} {\bibfnamefont {V.}~\bibnamefont
  {Steinberg}},\ }\bibfield  {title} {\bibinfo {title} {Critical slowing down
  in polymer dynamics near the coil-stretch transition in elongation flow},\
  }\href@noop {} {\bibfield  {journal} {\bibinfo  {journal} {Phys. Rev. E}\
  }\textbf {\bibinfo {volume} {78}},\ \bibinfo {pages} {040801(R)} (\bibinfo
  {year} {2008})}\BibitemShut {NoStop}%
\bibitem [{\citenamefont {Watanabe}\ and\ \citenamefont {Gotoh}(2010)}]{wg10}%
  \BibitemOpen
  \bibfield  {author} {\bibinfo {author} {\bibfnamefont {T.}~\bibnamefont
  {Watanabe}}\ and\ \bibinfo {author} {\bibfnamefont {T.}~\bibnamefont
  {Gotoh}},\ }\bibfield  {title} {\bibinfo {title} {Coil-stretch transition in
  an ensemble of polymers in isotropic turbulence},\ }\href@noop {} {\bibfield
  {journal} {\bibinfo  {journal} {Phys. Rev. {E}}\ }\textbf {\bibinfo {volume}
  {81}},\ \bibinfo {pages} {066301} (\bibinfo {year} {2010})}\BibitemShut
  {NoStop}%
\bibitem [{\citenamefont {Turitsyn}\ \emph {et~al.}(2007)\citenamefont
  {Turitsyn}, \citenamefont {Chertkov}, \citenamefont {Chernyak},\ and\
  \citenamefont {Puliafito}}]{tccp07}%
  \BibitemOpen
  \bibfield  {author} {\bibinfo {author} {\bibfnamefont {K.}~\bibnamefont
  {Turitsyn}}, \bibinfo {author} {\bibfnamefont {M.}~\bibnamefont {Chertkov}},
  \bibinfo {author} {\bibfnamefont {V.~Y.}\ \bibnamefont {Chernyak}},\ and\
  \bibinfo {author} {\bibfnamefont {A.}~\bibnamefont {Puliafito}},\ }\bibfield
  {title} {\bibinfo {title} {Statistics of entropy production in linearized
  stochastic systems},\ }\href@noop {} {\bibfield  {journal} {\bibinfo
  {journal} {Phys. Rev. Lett.}\ }\textbf {\bibinfo {volume} {98}},\ \bibinfo
  {pages} {180603} (\bibinfo {year} {2007})}\BibitemShut {NoStop}%
\bibitem [{\citenamefont {Latinwo}\ and\ \citenamefont
  {Schroeder}(2013)}]{ls13}%
  \BibitemOpen
  \bibfield  {author} {\bibinfo {author} {\bibfnamefont {F.}~\bibnamefont
  {Latinwo}}\ and\ \bibinfo {author} {\bibfnamefont {C.~M.}\ \bibnamefont
  {Schroeder}},\ }\bibfield  {title} {\bibinfo {title} {Nonequilibrium work
  relations for polymer dynamics in dilute solutions},\ }\href@noop {}
  {\bibfield  {journal} {\bibinfo  {journal} {Macromolecules}\ }\textbf
  {\bibinfo {volume} {46}},\ \bibinfo {pages} {8345} (\bibinfo {year}
  {2013})}\BibitemShut {NoStop}%
\bibitem [{\citenamefont {Latinwo}\ and\ \citenamefont
  {Schroeder}(2014)}]{latinwo14}%
  \BibitemOpen
  \bibfield  {author} {\bibinfo {author} {\bibfnamefont {F.}~\bibnamefont
  {Latinwo}}\ and\ \bibinfo {author} {\bibfnamefont {C.~M.}\ \bibnamefont
  {Schroeder}},\ }\bibfield  {title} {\bibinfo {title} {Determining elasticity
  from single polymer dynamics},\ }\href@noop {} {\bibfield  {journal}
  {\bibinfo  {journal} {Soft Matter}\ }\textbf {\bibinfo {volume} {10}},\
  \bibinfo {pages} {2178} (\bibinfo {year} {2014})}\BibitemShut {NoStop}%
\bibitem [{\citenamefont {Latinwo}\ \emph {et~al.}(2014)\citenamefont
  {Latinwo}, \citenamefont {Hsiao},\ and\ \citenamefont
  {Schroeder}}]{latinwo2014nonequilibrium}%
  \BibitemOpen
  \bibfield  {author} {\bibinfo {author} {\bibfnamefont {F.}~\bibnamefont
  {Latinwo}}, \bibinfo {author} {\bibfnamefont {K.-W.}\ \bibnamefont {Hsiao}},\
  and\ \bibinfo {author} {\bibfnamefont {C.~M.}\ \bibnamefont {Schroeder}},\
  }\bibfield  {title} {\bibinfo {title} {Nonequilibrium thermodynamics of
  dilute polymer solutions in flow},\ }\href@noop {} {\bibfield  {journal}
  {\bibinfo  {journal} {J. Chem. Phys.}\ }\textbf {\bibinfo {volume} {141}},\
  \bibinfo {pages} {174903} (\bibinfo {year} {2014})}\BibitemShut {NoStop}%
\bibitem [{\citenamefont {Vucelja}\ \emph {et~al.}(2015)\citenamefont
  {Vucelja}, \citenamefont {Turitsyn},\ and\ \citenamefont {Chertkov}}]{vtc15}%
  \BibitemOpen
  \bibfield  {author} {\bibinfo {author} {\bibfnamefont {M.}~\bibnamefont
  {Vucelja}}, \bibinfo {author} {\bibfnamefont {K.~S.}\ \bibnamefont
  {Turitsyn}},\ and\ \bibinfo {author} {\bibfnamefont {M.}~\bibnamefont
  {Chertkov}},\ }\bibfield  {title} {\bibinfo {title} {Extreme-value statistics
  of work done in stretching a polymer in a gradient flow},\ }\href@noop {}
  {\bibfield  {journal} {\bibinfo  {journal} {Phys. Rev. E}\ }\textbf {\bibinfo
  {volume} {91}},\ \bibinfo {pages} {022123} (\bibinfo {year}
  {2015})}\BibitemShut {NoStop}%
\bibitem [{\citenamefont {Ghosal}\ and\ \citenamefont {Cherayil}(2016)}]{gc16}%
  \BibitemOpen
  \bibfield  {author} {\bibinfo {author} {\bibfnamefont {A.}~\bibnamefont
  {Ghosal}}\ and\ \bibinfo {author} {\bibfnamefont {B.~J.}\ \bibnamefont
  {Cherayil}},\ }\bibfield  {title} {\bibinfo {title} {Polymer extension under
  flow: Some statistical properties of the work distribution function},\
  }\href@noop {} {\bibfield  {journal} {\bibinfo  {journal} {J. Chem. Phys.}\
  }\textbf {\bibinfo {volume} {145}},\ \bibinfo {pages} {204901} (\bibinfo
  {year} {2016})}\BibitemShut {NoStop}%
\bibitem [{\citenamefont {Ghosal}\ and\ \citenamefont {Cherayil}(2018)}]{gc18}%
  \BibitemOpen
  \bibfield  {author} {\bibinfo {author} {\bibfnamefont {A.}~\bibnamefont
  {Ghosal}}\ and\ \bibinfo {author} {\bibfnamefont {B.~J.}\ \bibnamefont
  {Cherayil}},\ }\bibfield  {title} {\bibinfo {title} {Anomalies in the
  coil-stretch transition of flexible polymers},\ }\href@noop {} {\bibfield
  {journal} {\bibinfo  {journal} {J. Chem. Phys.}\ }\textbf {\bibinfo {volume}
  {148}},\ \bibinfo {pages} {094903} (\bibinfo {year} {2018})}\BibitemShut
  {NoStop}%
\bibitem [{\citenamefont {Sultanov}\ \emph {et~al.}(2021)\citenamefont
  {Sultanov}, \citenamefont {Sultanova}, \citenamefont {Falkovich},
  \citenamefont {Lebedev}, \citenamefont {Liu},\ and\ \citenamefont
  {Steinberg}}]{ssflls21}%
  \BibitemOpen
  \bibfield  {author} {\bibinfo {author} {\bibfnamefont {F.}~\bibnamefont
  {Sultanov}}, \bibinfo {author} {\bibfnamefont {M.}~\bibnamefont {Sultanova}},
  \bibinfo {author} {\bibfnamefont {G.}~\bibnamefont {Falkovich}}, \bibinfo
  {author} {\bibfnamefont {V.}~\bibnamefont {Lebedev}}, \bibinfo {author}
  {\bibfnamefont {Y.}~\bibnamefont {Liu}},\ and\ \bibinfo {author}
  {\bibfnamefont {V.}~\bibnamefont {Steinberg}},\ }\bibfield  {title} {\bibinfo
  {title} {Entropic characterization of the coil-stretch transition of polymers
  in random flows},\ }\href@noop {} {\bibfield  {journal} {\bibinfo  {journal}
  {Phys. Rev. E}\ }\textbf {\bibinfo {volume} {103}},\ \bibinfo {pages}
  {033107} (\bibinfo {year} {2021})}\BibitemShut {NoStop}%
\bibitem [{\citenamefont {Groisman}\ and\ \citenamefont
  {Steinberg}(2000)}]{gs00}%
  \BibitemOpen
  \bibfield  {author} {\bibinfo {author} {\bibfnamefont {A.}~\bibnamefont
  {Groisman}}\ and\ \bibinfo {author} {\bibfnamefont {V.}~\bibnamefont
  {Steinberg}},\ }\bibfield  {title} {\bibinfo {title} {Elastic turbulence in a
  polymer solution flow},\ }\href@noop {} {\bibfield  {journal} {\bibinfo
  {journal} {Nature}\ }\textbf {\bibinfo {volume} {405}},\ \bibinfo {pages}
  {53} (\bibinfo {year} {2000})}\BibitemShut {NoStop}%
\bibitem [{\citenamefont {Bird}\ \emph {et~al.}(1987)\citenamefont {Bird},
  \citenamefont {Curtiss}, \citenamefont {Armstrong},\ and\ \citenamefont
  {Hassager}}]{bird}%
  \BibitemOpen
  \bibfield  {author} {\bibinfo {author} {\bibfnamefont {R.~B.}\ \bibnamefont
  {Bird}}, \bibinfo {author} {\bibfnamefont {C.~F.}\ \bibnamefont {Curtiss}},
  \bibinfo {author} {\bibfnamefont {R.~C.}\ \bibnamefont {Armstrong}},\ and\
  \bibinfo {author} {\bibfnamefont {O.}~\bibnamefont {Hassager}},\ }\href@noop
  {} {\emph {\bibinfo {title} {Dynamics of Polymeric Liquids}}},\ Vol.~\bibinfo
  {volume} {2}\ (\bibinfo  {publisher} {Wiley},\ \bibinfo {year}
  {1987})\BibitemShut {NoStop}%
\bibitem [{\citenamefont {Larson}(1988)}]{l88}%
  \BibitemOpen
  \bibfield  {author} {\bibinfo {author} {\bibfnamefont {R.~G.}\ \bibnamefont
  {Larson}},\ }\href@noop {} {\emph {\bibinfo {title} {Constitutive Equations
  for Polymer Melts and Solutions}}}\ (\bibinfo  {publisher} {Butterworth
  Publishers},\ \bibinfo {address} {Stoneham, MA},\ \bibinfo {year}
  {1988})\BibitemShut {NoStop}%
\bibitem [{\citenamefont {Graham}(2018)}]{g18}%
  \BibitemOpen
  \bibfield  {author} {\bibinfo {author} {\bibfnamefont {M.~D.}\ \bibnamefont
  {Graham}},\ }\href@noop {} {\emph {\bibinfo {title} {Microhydrodynamics,
  Brownian Motion, and Complex Fluids}}}\ (\bibinfo  {publisher} {Cambridge
  University Press},\ \bibinfo {address} {Cambridge, UK},\ \bibinfo {year}
  {2018})\BibitemShut {NoStop}%
\bibitem [{\citenamefont {Larson}\ \emph {et~al.}(1997)\citenamefont {Larson},
  \citenamefont {Perkins}, \citenamefont {Smith},\ and\ \citenamefont
  {Chu}}]{lpsc97}%
  \BibitemOpen
  \bibfield  {author} {\bibinfo {author} {\bibfnamefont {R.~G.}\ \bibnamefont
  {Larson}}, \bibinfo {author} {\bibfnamefont {T.~T.}\ \bibnamefont {Perkins}},
  \bibinfo {author} {\bibfnamefont {D.~E.}\ \bibnamefont {Smith}},\ and\
  \bibinfo {author} {\bibfnamefont {S.}~\bibnamefont {Chu}},\ }\bibfield
  {title} {\bibinfo {title} {Hydrodynamics of a {DNA} molecule in a flow
  field},\ }\href@noop {} {\bibfield  {journal} {\bibinfo  {journal} {Phys.
  Rev. E}\ }\textbf {\bibinfo {volume} {55}},\ \bibinfo {pages} {1794}
  (\bibinfo {year} {1997})}\BibitemShut {NoStop}%
\bibitem [{\citenamefont {Lumley}(1973)}]{l73}%
  \BibitemOpen
  \bibfield  {author} {\bibinfo {author} {\bibfnamefont {J.~L.}\ \bibnamefont
  {Lumley}},\ }\bibfield  {title} {\bibinfo {title} {Drag reduction in
  turbulent flow by polymer additives},\ }\href@noop {} {\bibfield  {journal}
  {\bibinfo  {journal} {J. Polymer Sci. Macromol. Rev.}\ }\textbf {\bibinfo
  {volume} {7}},\ \bibinfo {pages} {263} (\bibinfo {year} {1973})}\BibitemShut
  {NoStop}%
\bibitem [{\citenamefont {Balkovsky}\ \emph {et~al.}(2000)\citenamefont
  {Balkovsky}, \citenamefont {Fouxon},\ and\ \citenamefont {Lebedev}}]{bfl00}%
  \BibitemOpen
  \bibfield  {author} {\bibinfo {author} {\bibfnamefont {E.}~\bibnamefont
  {Balkovsky}}, \bibinfo {author} {\bibfnamefont {A.}~\bibnamefont {Fouxon}},\
  and\ \bibinfo {author} {\bibfnamefont {V.}~\bibnamefont {Lebedev}},\
  }\bibfield  {title} {\bibinfo {title} {Turbulent dynamics of polymer
  solutions},\ }\href@noop {} {\bibfield  {journal} {\bibinfo  {journal} {Phys.
  Rev. Lett.}\ }\textbf {\bibinfo {volume} {84}},\ \bibinfo {pages} {4765}
  (\bibinfo {year} {2000})}\BibitemShut {NoStop}%
\bibitem [{\citenamefont {Chertkov}(2000)}]{c00}%
  \BibitemOpen
  \bibfield  {author} {\bibinfo {author} {\bibfnamefont {M.}~\bibnamefont
  {Chertkov}},\ }\bibfield  {title} {\bibinfo {title} {Polymer stretching by
  turbulence},\ }\href@noop {} {\bibfield  {journal} {\bibinfo  {journal}
  {Phys. Rev. Lett.}\ }\textbf {\bibinfo {volume} {84}},\ \bibinfo {pages}
  {4761} (\bibinfo {year} {2000})}\BibitemShut {NoStop}%
\bibitem [{\citenamefont {Smith}\ \emph {et~al.}(1999)\citenamefont {Smith},
  \citenamefont {Babcock},\ and\ \citenamefont {Chu}}]{sbc99}%
  \BibitemOpen
  \bibfield  {author} {\bibinfo {author} {\bibfnamefont {D.~E.}\ \bibnamefont
  {Smith}}, \bibinfo {author} {\bibfnamefont {H.~P.}\ \bibnamefont {Babcock}},\
  and\ \bibinfo {author} {\bibfnamefont {S.}~\bibnamefont {Chu}},\ }\bibfield
  {title} {\bibinfo {title} {Single-polymer dynamics in steady shear flow},\
  }\href@noop {} {\bibfield  {journal} {\bibinfo  {journal} {Science}\ }\textbf
  {\bibinfo {volume} {283}},\ \bibinfo {pages} {1724} (\bibinfo {year}
  {1999})}\BibitemShut {NoStop}%
\bibitem [{\citenamefont {Schroeder}\ \emph
  {et~al.}(2005{\natexlab{a}})\citenamefont {Schroeder}, \citenamefont
  {Teixeira}, \citenamefont {Shaqfeh},\ and\ \citenamefont {Chu}}]{stsc05a}%
  \BibitemOpen
  \bibfield  {author} {\bibinfo {author} {\bibfnamefont {C.~M.}\ \bibnamefont
  {Schroeder}}, \bibinfo {author} {\bibfnamefont {R.~E.}\ \bibnamefont
  {Teixeira}}, \bibinfo {author} {\bibfnamefont {E.~S.~G.}\ \bibnamefont
  {Shaqfeh}},\ and\ \bibinfo {author} {\bibfnamefont {S.}~\bibnamefont {Chu}},\
  }\bibfield  {title} {\bibinfo {title} {Characteristic periodic motion of
  polymers in shear flow},\ }\href@noop {} {\bibfield  {journal} {\bibinfo
  {journal} {Phys. Rev. Lett.}\ }\textbf {\bibinfo {volume} {95}},\ \bibinfo
  {pages} {018301} (\bibinfo {year} {2005}{\natexlab{a}})}\BibitemShut
  {NoStop}%
\bibitem [{\citenamefont {Teixeira}\ \emph {et~al.}(2005)\citenamefont
  {Teixeira}, \citenamefont {Babcock}, \citenamefont {Shaqfeh},\ and\
  \citenamefont {Chu}}]{tbsc05}%
  \BibitemOpen
  \bibfield  {author} {\bibinfo {author} {\bibfnamefont {R.~E.}\ \bibnamefont
  {Teixeira}}, \bibinfo {author} {\bibfnamefont {H.~O.}\ \bibnamefont
  {Babcock}}, \bibinfo {author} {\bibfnamefont {E.~S.~G.}\ \bibnamefont
  {Shaqfeh}},\ and\ \bibinfo {author} {\bibfnamefont {S.}~\bibnamefont {Chu}},\
  }\bibfield  {title} {\bibinfo {title} {Shear thinning and tumbling dynamics
  of single polymers in the flow-gradient plane},\ }\href@noop {} {\bibfield
  {journal} {\bibinfo  {journal} {Macromolecules}\ }\textbf {\bibinfo {volume}
  {38}},\ \bibinfo {pages} {581} (\bibinfo {year} {2005})}\BibitemShut
  {NoStop}%
\bibitem [{\citenamefont {Gerashchenko}\ and\ \citenamefont
  {Steinberg}(2006)}]{gs06}%
  \BibitemOpen
  \bibfield  {author} {\bibinfo {author} {\bibfnamefont {S.}~\bibnamefont
  {Gerashchenko}}\ and\ \bibinfo {author} {\bibfnamefont {V.}~\bibnamefont
  {Steinberg}},\ }\bibfield  {title} {\bibinfo {title} {Statistics of tumbling
  of a single polymer molecule in shear flow},\ }\href@noop {} {\bibfield
  {journal} {\bibinfo  {journal} {Phys. Rev. Lett.}\ }\textbf {\bibinfo
  {volume} {96}},\ \bibinfo {pages} {038304} (\bibinfo {year}
  {2006})}\BibitemShut {NoStop}%
\bibitem [{\citenamefont {\"Ottinger}(1996)}]{o96}%
  \BibitemOpen
  \bibfield  {author} {\bibinfo {author} {\bibfnamefont {H.~C.}\ \bibnamefont
  {\"Ottinger}},\ }\href@noop {} {\emph {\bibinfo {title} {Stochastic Processes
  in Polymeric Fluids}}}\ (\bibinfo  {publisher} {Springer},\ \bibinfo
  {address} {Berlin},\ \bibinfo {year} {1996})\BibitemShut {NoStop}%
\bibitem [{\citenamefont {Falkovich}\ \emph {et~al.}(2001)\citenamefont
  {Falkovich}, \citenamefont {Gaw\c{e}dzki},\ and\ \citenamefont
  {Vergassola}}]{fgv01}%
  \BibitemOpen
  \bibfield  {author} {\bibinfo {author} {\bibfnamefont {G.}~\bibnamefont
  {Falkovich}}, \bibinfo {author} {\bibfnamefont {K.}~\bibnamefont
  {Gaw\c{e}dzki}},\ and\ \bibinfo {author} {\bibfnamefont {M.}~\bibnamefont
  {Vergassola}},\ }\bibfield  {title} {\bibinfo {title} {Particles and fields
  in fluid turbulence},\ }\href@noop {} {\bibfield  {journal} {\bibinfo
  {journal} {Rev. Mod. Phys.}\ }\textbf {\bibinfo {volume} {73}},\ \bibinfo
  {pages} {913} (\bibinfo {year} {2001})}\BibitemShut {NoStop}%
\bibitem [{\citenamefont {{E.~L.~C.~VI~M. Plan}}\ \emph
  {et~al.}(2016)\citenamefont {{E.~L.~C.~VI~M. Plan}}, \citenamefont {Ali},\
  and\ \citenamefont {Vincenzi}}]{pav16}%
  \BibitemOpen
  \bibfield  {author} {\bibinfo {author} {\bibnamefont {{E.~L.~C.~VI~M.
  Plan}}}, \bibinfo {author} {\bibfnamefont {A.}~\bibnamefont {Ali}},\ and\
  \bibinfo {author} {\bibfnamefont {D.}~\bibnamefont {Vincenzi}},\ }\bibfield
  {title} {\bibinfo {title} {Bead-rod-spring models in random flows},\
  }\href@noop {} {\bibfield  {journal} {\bibinfo  {journal} {Phys. Rev. E}\
  }\textbf {\bibinfo {volume} {94}},\ \bibinfo {pages} {020501(R)} (\bibinfo
  {year} {2016})}\BibitemShut {NoStop}%
\bibitem [{\citenamefont {{Martins Afonso}}\ and\ \citenamefont
  {Vincenzi}(2005)}]{mav05}%
  \BibitemOpen
  \bibfield  {author} {\bibinfo {author} {\bibfnamefont {M.}~\bibnamefont
  {{Martins Afonso}}}\ and\ \bibinfo {author} {\bibfnamefont {D.}~\bibnamefont
  {Vincenzi}},\ }\bibfield  {title} {\bibinfo {title} {Nonlinear elastic
  polymers in random flow},\ }\href@noop {} {\bibfield  {journal} {\bibinfo
  {journal} {J. Fluid Mech.}\ }\textbf {\bibinfo {volume} {540}},\ \bibinfo
  {pages} {99} (\bibinfo {year} {2005})}\BibitemShut {NoStop}%
\bibitem [{\citenamefont {James}\ and\ \citenamefont {Ray}(2017)}]{jr17}%
  \BibitemOpen
  \bibfield  {author} {\bibinfo {author} {\bibfnamefont {M.}~\bibnamefont
  {James}}\ and\ \bibinfo {author} {\bibfnamefont {S.~S.}\ \bibnamefont
  {Ray}},\ }\bibfield  {title} {\bibinfo {title} {Enhanced droplet collision
  rates and impact velocities in turbulent flows: The effect of poly-dispersity
  and transient phases},\ }\href@noop {} {\bibfield  {journal} {\bibinfo
  {journal} {Sci. Reports}\ }\textbf {\bibinfo {volume} {7}},\ \bibinfo {pages}
  {12231} (\bibinfo {year} {2017})}\BibitemShut {NoStop}%
\bibitem [{\citenamefont {Vincenzi}\ \emph {et~al.}(2021)\citenamefont
  {Vincenzi}, \citenamefont {Watanabe}, \citenamefont {Ray},\ and\
  \citenamefont {Picardo}}]{vwrp21}%
  \BibitemOpen
  \bibfield  {author} {\bibinfo {author} {\bibfnamefont {D.}~\bibnamefont
  {Vincenzi}}, \bibinfo {author} {\bibfnamefont {T.}~\bibnamefont {Watanabe}},
  \bibinfo {author} {\bibfnamefont {S.~S.}\ \bibnamefont {Ray}},\ and\ \bibinfo
  {author} {\bibfnamefont {J.}~\bibnamefont {Picardo}},\ }\bibfield  {title}
  {\bibinfo {title} {Polymer scission in turbulent flows},\ }\href@noop {}
  {\bibfield  {journal} {\bibinfo  {journal} {J. Fluid Mech.}\ }\textbf
  {\bibinfo {volume} {912}},\ \bibinfo {pages} {A18} (\bibinfo {year}
  {2021})}\BibitemShut {NoStop}%
\bibitem [{\citenamefont {Bec}\ \emph {et~al.}(2006)\citenamefont {Bec},
  \citenamefont {Biferale}, \citenamefont {Boffetta}, \citenamefont {Cencini},
  \citenamefont {o},\ and\ \citenamefont {Toschi}}]{bbbcmt06}%
  \BibitemOpen
  \bibfield  {author} {\bibinfo {author} {\bibfnamefont {J.}~\bibnamefont
  {Bec}}, \bibinfo {author} {\bibfnamefont {L.}~\bibnamefont {Biferale}},
  \bibinfo {author} {\bibfnamefont {G.}~\bibnamefont {Boffetta}}, \bibinfo
  {author} {\bibfnamefont {M.}~\bibnamefont {Cencini}}, \bibinfo {author}
  {\bibfnamefont {S.~M.}\ \bibnamefont {o}},\ and\ \bibinfo {author}
  {\bibfnamefont {F.}~\bibnamefont {Toschi}},\ }\bibfield  {title} {\bibinfo
  {title} {{Lyapunov} exponents of heavy particles in turbulence},\ }\href@noop
  {} {\bibfield  {journal} {\bibinfo  {journal} {Phys. Fluids}\ }\textbf
  {\bibinfo {volume} {18}},\ \bibinfo {pages} {091702} (\bibinfo {year}
  {2006})}\BibitemShut {NoStop}%
\bibitem [{\citenamefont {Boffetta}\ and\ \citenamefont
  {Musacchio}(2017)}]{bm17}%
  \BibitemOpen
  \bibfield  {author} {\bibinfo {author} {\bibfnamefont {G.}~\bibnamefont
  {Boffetta}}\ and\ \bibinfo {author} {\bibfnamefont {S.}~\bibnamefont
  {Musacchio}},\ }\bibfield  {title} {\bibinfo {title} {Chaos and
  predictability of homogeneous-isotropic turbulence},\ }\href@noop {}
  {\bibfield  {journal} {\bibinfo  {journal} {Phys. Rev. Lett.}\ }\textbf
  {\bibinfo {volume} {119}},\ \bibinfo {pages} {054102} (\bibinfo {year}
  {2017})}\BibitemShut {NoStop}%
\bibitem [{\citenamefont {Boffetta}\ \emph {et~al.}(2003)\citenamefont
  {Boffetta}, \citenamefont {Celani},\ and\ \citenamefont {Musacchio}}]{bcm03}%
  \BibitemOpen
  \bibfield  {author} {\bibinfo {author} {\bibfnamefont {G.}~\bibnamefont
  {Boffetta}}, \bibinfo {author} {\bibfnamefont {A.}~\bibnamefont {Celani}},\
  and\ \bibinfo {author} {\bibfnamefont {S.}~\bibnamefont {Musacchio}},\
  }\bibfield  {title} {\bibinfo {title} {Two-dimensional turbulence of dilute
  polymer solutions},\ }\href@noop {} {\bibfield  {journal} {\bibinfo
  {journal} {Phys. Rev. Lett.}\ }\textbf {\bibinfo {volume} {91}},\ \bibinfo
  {pages} {034501} (\bibinfo {year} {2003})}\BibitemShut {NoStop}%
\bibitem [{\citenamefont {Ford}(1963)}]{Jaynes63}%
  \BibitemOpen
  \bibfield  {author} {\bibinfo {author} {\bibfnamefont {K.~W.}\ \bibnamefont
  {Ford}},\ }\bibinfo {title} {Brandeis lectures in statistical physics}\
  (\bibinfo  {publisher} {W.A. Benjamin Inc.},\ \bibinfo {address} {New York},\
  \bibinfo {year} {1963})\ Chap.~\bibinfo {chapter} {4}\BibitemShut {NoStop}%
\bibitem [{\citenamefont {Kullback}\ and\ \citenamefont
  {Leibler}(1951)}]{kullback195110}%
  \BibitemOpen
  \bibfield  {author} {\bibinfo {author} {\bibfnamefont {S.}~\bibnamefont
  {Kullback}}\ and\ \bibinfo {author} {\bibfnamefont {R.~A.}\ \bibnamefont
  {Leibler}},\ }\bibfield  {title} {\bibinfo {title} {On information and
  sufficiency},\ }\href@noop {} {\bibfield  {journal} {\bibinfo  {journal}
  {Ann. Math. Stat}\ }\textbf {\bibinfo {volume} {22}},\ \bibinfo {pages} {79}
  (\bibinfo {year} {1951})}\BibitemShut {NoStop}%
\bibitem [{\citenamefont {Perkins}\ \emph {et~al.}(1999)\citenamefont
  {Perkins}, \citenamefont {Smith},\ and\ \citenamefont {Chu}}]{psc99}%
  \BibitemOpen
  \bibfield  {author} {\bibinfo {author} {\bibfnamefont {T.~T.}\ \bibnamefont
  {Perkins}}, \bibinfo {author} {\bibfnamefont {D.~E.}\ \bibnamefont {Smith}},\
  and\ \bibinfo {author} {\bibfnamefont {S.}~\bibnamefont {Chu}},\ }\bibfield
  {title} {\bibinfo {title} {Single polymers in elongational flows: Dynamic,
  steady-state, and population-averaged properties},\ }in\ \href@noop {} {\emph
  {\bibinfo {booktitle} {Flexible Polymer Chains in Elongational Flow}}},\
  \bibinfo {editor} {edited by\ \bibinfo {editor} {\bibfnamefont {T.~Q.}\
  \bibnamefont {Nguyen}}\ and\ \bibinfo {editor} {\bibfnamefont {H.-H.}\
  \bibnamefont {Kausch}}}\ (\bibinfo  {publisher} {Springer},\ \bibinfo
  {address} {Berlin, Heidelberg},\ \bibinfo {year} {1999})\ pp.\ \bibinfo
  {pages} {283--334}\BibitemShut {NoStop}%
\bibitem [{\citenamefont {Schroeder}\ \emph
  {et~al.}(2005{\natexlab{b}})\citenamefont {Schroeder}, \citenamefont
  {Teixeira}, \citenamefont {Shaqfeh},\ and\ \citenamefont {Chu}}]{stsc05b}%
  \BibitemOpen
  \bibfield  {author} {\bibinfo {author} {\bibfnamefont {C.~M.}\ \bibnamefont
  {Schroeder}}, \bibinfo {author} {\bibfnamefont {R.~E.}\ \bibnamefont
  {Teixeira}}, \bibinfo {author} {\bibfnamefont {E.~S.~G.}\ \bibnamefont
  {Shaqfeh}},\ and\ \bibinfo {author} {\bibfnamefont {S.}~\bibnamefont {Chu}},\
  }\bibfield  {title} {\bibinfo {title} {Dynamics of {DNA} in the flow-gradient
  plane of steady shear flow: Observations and simulations},\ }\href@noop {}
  {\bibfield  {journal} {\bibinfo  {journal} {Macromolecules}\ }\textbf
  {\bibinfo {volume} {38}},\ \bibinfo {pages} {1967} (\bibinfo {year}
  {2005}{\natexlab{b}})}\BibitemShut {NoStop}%
\bibitem [{\citenamefont {Celani}\ \emph {et~al.}(2005)\citenamefont {Celani},
  \citenamefont {Puliafito},\ and\ \citenamefont {Turitsyn}}]{cpt05}%
  \BibitemOpen
  \bibfield  {author} {\bibinfo {author} {\bibfnamefont {A.}~\bibnamefont
  {Celani}}, \bibinfo {author} {\bibfnamefont {A.}~\bibnamefont {Puliafito}},\
  and\ \bibinfo {author} {\bibfnamefont {K.}~\bibnamefont {Turitsyn}},\
  }\bibfield  {title} {\bibinfo {title} {Polymers in linear shear flow: A
  numerical study},\ }\href@noop {} {\bibfield  {journal} {\bibinfo  {journal}
  {Europhys. Lett.}\ }\textbf {\bibinfo {volume} {70}},\ \bibinfo {pages} {464}
  (\bibinfo {year} {2005})}\BibitemShut {NoStop}%
\bibitem [{\citenamefont {Musacchio}\ and\ \citenamefont
  {Vincenzi}(2011)}]{mv11}%
  \BibitemOpen
  \bibfield  {author} {\bibinfo {author} {\bibfnamefont {S.}~\bibnamefont
  {Musacchio}}\ and\ \bibinfo {author} {\bibfnamefont {D.}~\bibnamefont
  {Vincenzi}},\ }\bibfield  {title} {\bibinfo {title} {Deformation of a
  flexible polymer in a random flow with long correlation time},\ }\href@noop
  {} {\bibfield  {journal} {\bibinfo  {journal} {J. Fluid Mech.}\ }\textbf
  {\bibinfo {volume} {670}},\ \bibinfo {pages} {326} (\bibinfo {year}
  {2011})}\BibitemShut {NoStop}%
\bibitem [{\citenamefont {Risken}(1989)}]{r89}%
  \BibitemOpen
  \bibfield  {author} {\bibinfo {author} {\bibfnamefont {H.}~\bibnamefont
  {Risken}},\ }\href@noop {} {\emph {\bibinfo {title} {The Fokker--Planck
  Equation}}}\ (\bibinfo  {publisher} {Springer},\ \bibinfo {address}
  {Berlin},\ \bibinfo {year} {1989})\BibitemShut {NoStop}%
\bibitem [{\citenamefont {Salazar}\ and\ \citenamefont {Collins}(2009)}]{sc09}%
  \BibitemOpen
  \bibfield  {author} {\bibinfo {author} {\bibfnamefont {J.~P. L.~C.}\
  \bibnamefont {Salazar}}\ and\ \bibinfo {author} {\bibfnamefont {L.~R.}\
  \bibnamefont {Collins}},\ }\bibfield  {title} {\bibinfo {title} {Two-particle
  dispersion in isotropic turbulent flows},\ }\href@noop {} {\bibfield
  {journal} {\bibinfo  {journal} {Annu. Rev. Fluid Mech.}\ }\textbf {\bibinfo
  {volume} {41}},\ \bibinfo {pages} {405} (\bibinfo {year} {2009})}\BibitemShut
  {NoStop}%
\bibitem [{\citenamefont {Vaithianathan}\ and\ \citenamefont
  {Collins}(2003)}]{vc03}%
  \BibitemOpen
  \bibfield  {author} {\bibinfo {author} {\bibfnamefont {T.}~\bibnamefont
  {Vaithianathan}}\ and\ \bibinfo {author} {\bibfnamefont {L.~R.}\ \bibnamefont
  {Collins}},\ }\bibfield  {title} {\bibinfo {title} {Numerical approach to
  simulating turbulent flow of a viscoelastic polymer solution},\ }\href@noop
  {} {\bibfield  {journal} {\bibinfo  {journal} {J. Comp. Phys.}\ }\textbf
  {\bibinfo {volume} {187}},\ \bibinfo {pages} {1} (\bibinfo {year}
  {2003})}\BibitemShut {NoStop}%
\bibitem [{\citenamefont {Fattal}\ and\ \citenamefont
  {Kupferman}(2004)}]{fk04}%
  \BibitemOpen
  \bibfield  {author} {\bibinfo {author} {\bibfnamefont {R.}~\bibnamefont
  {Fattal}}\ and\ \bibinfo {author} {\bibfnamefont {R.}~\bibnamefont
  {Kupferman}},\ }\bibfield  {title} {\bibinfo {title} {Constitutive laws for
  the matrix-logarithm of the conformation tensor},\ }\href@noop {} {\bibfield
  {journal} {\bibinfo  {journal} {J. Non-Newtonian Fluid Mech.}\ }\textbf
  {\bibinfo {volume} {123}},\ \bibinfo {pages} {281} (\bibinfo {year}
  {2004})}\BibitemShut {NoStop}%
\bibitem [{\citenamefont {Hameduddin}\ \emph {et~al.}(2018)\citenamefont
  {Hameduddin}, \citenamefont {Meneveau}, \citenamefont {Zaki},\ and\
  \citenamefont {Gayme}}]{hmzg18}%
  \BibitemOpen
  \bibfield  {author} {\bibinfo {author} {\bibfnamefont {I.}~\bibnamefont
  {Hameduddin}}, \bibinfo {author} {\bibfnamefont {C.}~\bibnamefont
  {Meneveau}}, \bibinfo {author} {\bibfnamefont {T.~A.}\ \bibnamefont {Zaki}},\
  and\ \bibinfo {author} {\bibfnamefont {D.~F.}\ \bibnamefont {Gayme}},\
  }\bibfield  {title} {\bibinfo {title} {Geometric decomposition of the
  conformation tensor in viscoelastic turbulence},\ }\href@noop {} {\bibfield
  {journal} {\bibinfo  {journal} {J. Fluid Mech.}\ }\textbf {\bibinfo {volume}
  {842}},\ \bibinfo {pages} {395} (\bibinfo {year} {2018})}\BibitemShut
  {NoStop}%
\bibitem [{\citenamefont {Terrapon}\ \emph {et~al.}(2004)\citenamefont
  {Terrapon}, \citenamefont {Dubief}, \citenamefont {Moin}, \citenamefont
  {Shaqfeh},\ and\ \citenamefont {Lele}}]{terrapon}%
  \BibitemOpen
  \bibfield  {author} {\bibinfo {author} {\bibfnamefont {V.~E.}\ \bibnamefont
  {Terrapon}}, \bibinfo {author} {\bibfnamefont {Y.}~\bibnamefont {Dubief}},
  \bibinfo {author} {\bibfnamefont {P.}~\bibnamefont {Moin}}, \bibinfo {author}
  {\bibfnamefont {E.~S.~G.}\ \bibnamefont {Shaqfeh}},\ and\ \bibinfo {author}
  {\bibfnamefont {S.~K.}\ \bibnamefont {Lele}},\ }\bibfield  {title} {\bibinfo
  {title} {Simulated polymer stretch in a turbulent flow using {B}rownian
  dynamics},\ }\href@noop {} {\bibfield  {journal} {\bibinfo  {journal} {J.
  Fluid Mech.}\ }\textbf {\bibinfo {volume} {504}},\ \bibinfo {pages} {61}
  (\bibinfo {year} {2004})}\BibitemShut {NoStop}%
\bibitem [{\citenamefont {Peters}\ and\ \citenamefont
  {Schumacher}(2007)}]{ps07}%
  \BibitemOpen
  \bibfield  {author} {\bibinfo {author} {\bibfnamefont {T.}~\bibnamefont
  {Peters}}\ and\ \bibinfo {author} {\bibfnamefont {J.}~\bibnamefont
  {Schumacher}},\ }\bibfield  {title} {\bibinfo {title} {Two-way coupling of
  finitely extensible nonlinear elastic dumbbells with a turbulent shear
  flow},\ }\href@noop {} {\bibfield  {journal} {\bibinfo  {journal} {Phys.
  Fluids}\ }\textbf {\bibinfo {volume} {19}},\ \bibinfo {pages} {065109}
  (\bibinfo {year} {2007})}\BibitemShut {NoStop}%
\bibitem [{\citenamefont {Blackburn}\ \emph {et~al.}(1996)\citenamefont
  {Blackburn}, \citenamefont {Mansour},\ and\ \citenamefont
  {Cantwell}}]{bmc96}%
  \BibitemOpen
  \bibfield  {author} {\bibinfo {author} {\bibfnamefont {H.~M.}\ \bibnamefont
  {Blackburn}}, \bibinfo {author} {\bibfnamefont {N.~N.}\ \bibnamefont
  {Mansour}},\ and\ \bibinfo {author} {\bibfnamefont {B.~J.}\ \bibnamefont
  {Cantwell}},\ }\bibfield  {title} {\bibinfo {title} {Topology of fine-scale
  motions in turbulent channel flow},\ }\href@noop {} {\bibfield  {journal}
  {\bibinfo  {journal} {J. Fluid Mech.}\ }\textbf {\bibinfo {volume} {310}},\
  \bibinfo {pages} {269} (\bibinfo {year} {1996})}\BibitemShut {NoStop}%
\bibitem [{\citenamefont {Jin}\ and\ \citenamefont {Collins}(2007)}]{jc07}%
  \BibitemOpen
  \bibfield  {author} {\bibinfo {author} {\bibfnamefont {S.}~\bibnamefont
  {Jin}}\ and\ \bibinfo {author} {\bibfnamefont {L.~R.}\ \bibnamefont
  {Collins}},\ }\bibfield  {title} {\bibinfo {title} {Dynamics of dissolved
  polymer chains in isotropic turbulence},\ }\href@noop {} {\bibfield
  {journal} {\bibinfo  {journal} {New J. Phys.}\ }\textbf {\bibinfo {volume}
  {9}},\ \bibinfo {pages} {360} (\bibinfo {year} {2007})}\BibitemShut {NoStop}%
\end{thebibliography}%

\end{document}